\documentclass{aa}  

\usepackage{graphicx}
\usepackage{amsmath}
\usepackage{booktabs}
\usepackage{listings}
\usepackage{subcaption}
\usepackage{placeins}

\usepackage[pdftex,
            colorlinks=true,
            linkcolor=blue,
            citecolor=blue,
            urlcolor=blue,
            hypertexnames=false]{hyperref}

\renewcommand{\linenumberfont}{\relax}
\renewcommand{\thelinenumber}{}

\begin{document}

\title{Photometric classification of quasars from DES and photo-z estimation with Machine Learning}

\subtitle{}

\author{Pablo Motta\inst{1,2,3}\corrauth{pablomotta@ustc.edu.cn}
        \and Filipe B. Abdalla\inst{1,2,4}\corrauth{filipe.abdalla@gmail.com}
        \and Elcio Abdalla\inst{5,6}\corrauth{eabdalla@usp.br}
        \and Gabriel S. Costa\inst{7,3}
        \and Camila Cardoso\inst{8}
}

\institute{Department of Astronomy, School of Physical Sciences, University of Science and Technology of China, Hefei, Anhui 230026, China
        \and School of Astronomy and Space Science, University of Science and Technology of China, Hefei, Anhui 230026, China 
        \and Instituto de F\'isica, Universidade de S\~ao Paulo - C.P. 66318, CEP: 05315-970, S\~ao Paulo, Brazil
        \and Department of Physics and Electronics, Rhodes University, PO Box 94, Grahamstown, 6140, South Africa 
        \and Departamento de Física, Centro de Ciências Exatas e da Natureza, Universidade Federal da Paraíba, CEP 58059-970, João Pessoa, Brazil 
        \and Secretaria de Estado da Ciência, Tecnologia, Inovação e Ensino Superior, Governo da Paraíba, Brazil
        \and Instituto Nacional de Pesquisas Espaciais, Divisão de Astrof\'isica, Av. dos Astronautas, 1758, 12227-010 - São Jos\'e dos Campos, SP, Brazil
        \and Escola de Engenharia de Lorena (EEL), Universidade de São Paulo (USP), Estrada Municipal do Campinho, s/n, Lorena, CEP 12602-810, Brazil 
}

\date{Accepted XXX. Received YYY; in original form ZZZ}

% Abstract with 5 mandatory fields
\abstract
% Context
{Modern astronomy leverages photometric data from large surveys like the Dark Energy Survey (DES), LSST, and Euclid. Among billions of detected objects, quasars are invaluable probes of cosmic history. Their clustering provides key information on the large-scale matter distribution and baryon acoustic oscillations, while absorption features trace the intergalactic medium. However, their photometric similarity to stars often leads to misclassification and exclusion from cosmological analyses, undermining their utility for precision cosmology.}
% Aims
{To resolve this paradox and fully leverage quasars for cosmology, we construct a dedicated quasar catalog from photometric survey data. This effort requires addressing two requirements: (1) improving photometric classification to reliably distinguish quasars from stars, and (2) refining redshift estimation to accurately map their three-dimensional distribution.}
% Methods
{We perform a positional cross-match between photometric point-like objects from DES Data Release 2 (DES DR2) and spectroscopically classified objects from SDSS Data Release 16 (SDSS DR16). We employ a K-Nearest Neighbors (KNN) algorithm using PSF magnitudes in the $g$, $r$, $i$, and $z$ bands as features to classify quasars/galaxies against stellar contaminants. For photometric redshift estimation we use a hybrid machine learning approach combining a Boosted Decision Tree (BDT) from {\tt ANNz} and a Decision Tree Regressor (DTR) from scikit-learn. We further develop a stacked outlier classifier to mitigate catastrophic redshift errors.}
% Results
{Our KNN classification achieves a recall of 0.77 at 0.99 precision. The photometric redshift distribution spans $z \approx 0.5$ to $z > 3$, with a distinct, well-recovered population at $z \approx 4$. The full photometric redshift sample contains 872,372 objects; the reduced-outlier catalog contains 675,683 objects. The high-redshift sample ($z \gtrsim 3.5$) remains reliable without outlier removal as outliers do not extend beyond $z \approx 3.5$.}
% Conclusions
{We present a robustly characterized quasar catalog from DES DR2, suitable for cosmological studies. The cleaner catalog is ideal for large-scale structure probes in the $0<z<3$ range, while the full catalog can be used at $z \approx 4$. This catalog complements a galaxy-based catalog from DES, enabling joint cosmological analyses.}


\maketitle

\section{Introduction}
% Content of introduction from original paper
Modern astronomy leverages photometric data from large surveys like the Dark Energy Survey (DES; \citep{des2016, descollaboration2005, to2021dark, kessler2015results}), LSST \citep{ivezic2019lsst}, and Euclid \citep{laureijs2011euclid, Amendola_2013} to study the universe. Among billions of detected objects, quasars \citep{rees1984black} are invaluable probes of cosmic history \citep{van2010shear, fan2023quasars, zheng2020multiple, Frieman_2008}.
Their clustering provides key information on the large-scale matter distribution and baryon acoustic oscillations (BAO) \citep{eisenstein2005detection, Cole_2005, Anderson_2014, Alam_2017, huetsi2006acoustic}, while absorption features trace the intergalactic medium \citep{zhang2023trinity, rauch1998lyman}.
However, their photometric similarity to stars often leads to misclassification and exclusion from cosmological analyses \citep{repp2016systematic, richards2002color, tanaka2018photometric, richards2001photometric}, undermining their utility for precision cosmology \citep{joudaki2018kids, hildebrandt2017kids}.

To resolve this paradox and fully leverage quasars for cosmology, we propose constructing a dedicated quasar catalog from photometric survey data.
This effort requires addressing two requirements: (1) improving photometric classification to reliably distinguish quasars from stars, and (2) refining redshift estimation to accurately map their three-dimensional distribution.
These tasks are interdependent—misclassification propagates errors into redshift measurements \citep{hildebrandt2010photometric}, while precise redshifts help validate classifications.
Traditional methods, like color cuts \citep{richards2006efficiency} or morphology \citep{baldry2002morphological}, often fail to resolve ambiguities at high redshifts where quasar UV features shift into optical bands \citep{fan1999evolution, bolton2012spectral, fan1999high}.

Quasars truly act as cosmic lighthouses: their extreme brilliance allows detection across vast cosmic distances, making them invaluable for mapping the universe's large-scale structure and probing the intergalactic medium via absorption features like the Lyman-$\alpha$ forest \citep{rauch1998lyman}.
Despite their significance, their photometric identification remains challenging.
In astronomical images, quasars appear as unresolved points of light, morphologically indistinguishable from stars \citep{Weiner2005}.
Their colors, measured through broad photometric bands, often overlap with certain stellar types, such as hot blue stars \citep{Richards2002} or cool red dwarfs \citep{skrzypek2016}.
Misclassifying quasars introduces systematic errors into photometric redshifts, skewing cosmological measurements like the BAO scale \citep{eisenstein2005detection, Cole_2005, Anderson_2014, Alam_2017}—a fundamental ``ruler'' for studying the universe's expansion history \citep{prat2018dark}.
Thus, enhancing both quasar classification purity and photometric redshift precision are inseparable and equally crucial goals, forming the dual focus of this work.

Quasars are among the most energetic objects in the universe, fueled by supermassive black holes \citep{rees1984black}.
Their light spans the electromagnetic spectrum, exhibiting distinctive ultraviolet (UV) glows and broad emission lines \citep{vandenberk2001composite}.
However, in photometric surveys, their point-like appearance and characteristic colors frequently mimic those of stars \citep{richards2002color, brescia2013photometric}.
At higher redshifts, this confusion intensifies due to the cosmological redshift effect.
For instance, low-redshift quasars may resemble young, hot stars (A-type), while high-redshift quasars can be mistaken for cool, dim stars (M dwarfs) in broadband photometry \citep{fan2001color}.
Traditional identification methods, such as fixed color-based selection criteria \citep{richards2006efficiency} or morphological analysis \citep{baldry2002morphological}, struggle significantly to resolve these intrinsic ambiguities, especially in noisy data or surveys with limited wavelength coverage \citep{tanaka2018photometric}.
Critically, these persistent misclassifications propagate into photometric redshift errors \citep{hildebrandt2010photometric, wu2004color}, distorting three-dimensional maps of the universe's matter distribution and ultimately compromising cosmological conclusions \citep{abbott2019dark, PhysRevD.100.043501}.

To overcome these limitations, we turn to the advanced capabilities of machine learning (ML).
Unlike simplistic traditional techniques, ML algorithms can simultaneously analyze a multitude of features—such as brightness, colors, and subtle spatial patterns—to effectively disentangle quasars from stars while concurrently refining their redshift estimates \citep{hoyle2015photometric, beck2017photometric, rumelhart1986learning, bottou2010large, friedman2001greedy, Oyaizu_2008, almosallam2016gpz, almosallam2016sparse, rivera2018degradation}.
This powerful dual capability directly addresses the core challenge of modern cosmology: balancing immense statistical power from vast datasets with stringent precision requirements.

Photometric redshifts, estimated from broad-band colors rather than computationally intensive detailed spectra, offer an efficient means of determining distances for enormous datasets from surveys like DES, SDSS \citep{york2000sdss}, or Pan-STARRS \citep{kaiser2010panstarrs}.
However, their accuracy depends critically on the quality and representativeness of the training data \citep{hildebrandt2010photometric, newman2013deep2}.
For quasars, reliable redshift estimation necessitates a clean, comprehensive, and representative sample of spectroscopically confirmed quasars as ground truth \citep{dawson2013sdss, masters2019complete}.
In this study, we integrate spectroscopic quasar identifications from the Sloan Digital Sky Survey (SDSS)—spanning redshifts $z \approx 0.1$ to 6—with DES photometry to train our ML models.
Employing the Python library \texttt{scikit-learn} \citep{pedregosa2011scikit} and the public code {\tt ANNz} \citep{anzu2018}, we aim for two synergistic outcomes: high-purity quasar catalogs and precise photometric redshifts to create tomographic maps of the matter distribution.

Our work complements that presented in \cite{Abdalla2025machine}.
While both papers create DES catalogs for cosmological purposes, a key distinction lies in the type of celestial objects analyzed.
Their work focuses on extended objects (galaxies), whereas our study focuses on point-like objects, typically excluded from the photometric selection processes.
Consequently, this work not only focuses on accurate photo-z estimation but also properly separates galaxies and quasars from prevalent stellar contamination.
These two catalogs are complementary, and combining them would significantly enhance the overall cosmological performance of future analyses.

This complementarity opens new avenues for joint cosmological analyses, particularly in synergy with radio surveys targeting large-scale structure, such as CHIME \citep{bandura2014canadian}, FAST \citep{nan2011five, bigot2015hi}, SKA \citep{santos2015cosmology}, and Tianlai~\citep{chen2012tianlai}.
In particular, the Baryon Acoustic Oscillations from Integrated Neutral Gas Observations \citep[BINGO;][]{abdalla2022bingo, wuensche2022bingo, abdalla2022bingo-iii, liccardo2022bingo, fornazier2022bingo, zhang2022bingo, costa2022bingo, novaes2022bingo, santos2023bingo} experiment offers a unique opportunity to cross-correlate 21 cm intensity mapping data with optical catalogs.
A well-characterized optical catalog provides an ideal target for these cross-correlations.
Whether used individually or in combination, the galaxy and quasar catalogs can enhance tomographic analyses, mitigate systematics through cross-correlation, and improve constraints on cosmological parameters such as dark energy, structure growth, and primordial non-Gaussianity.
Therefore, these parallel efforts form a coherent framework that significantly boosts the scientific return of both optical and radio surveys.

This paper is organized as follows: in Section~\ref{sec.data} we describe the spectroscopic and photometric data; Section~\ref{sec.quasar_selection} details the optimization of ML models for classifying objects as stars or quasars and creating a quasar catalog; Section~\ref{sec.photo_z} describes photometric redshift estimation; and finally, in Section~\ref{sec.applying_on_des} we detail the application of our optimized classification and redshift estimation models to the full Dark Energy Survey data to create the comprehensive photometric catalog of DES quasars.

\section{Data}
\label{sec.data}

We performed a positional cross-match between photometric point-like objects from the Dark Energy Survey Data Release 2 \citep[DES DR2;][]{abbott2021dark} and spectroscopically classified objects from the Sloan Digital Sky Survey Data Release 16 \citep[SDSS DR16;][]{ahumada202016th}. Our specific interest is in identifying point-like sources, particularly quasars, which serve as valuable tracers for cosmological studies of large-scale structure and cosmic evolution. At the same time, it is essential to reliably identify and exclude stars, which act as contaminants in quasar samples used for cosmology. The cross-match allows us to associate the precise spectroscopic labels from SDSS—providing object type (quasar, star, galaxy) and redshift—with the photometric measurements from DES, enabling the construction of a training set for photometric classification and redshift estimation. In the following, we describe the two surveys used and detail the cross-matching procedure between them.

\subsection{The Dark Energy Survey}

The Dark Energy Survey \citep[DES;][]{abbott2018dark, abbott2021dark} is an optical and near-infrared public survey designed to improve our understanding of dark energy and cosmic acceleration by mapping galaxies, detecting supernovae, and studying cosmic structure patterns \cite{to2021dark}. The survey was conducted using the Dark Energy Camera (DECam) at the Cerro Tololo Inter-American Observatory (CTIO) and covers approximately 5000 deg$^2$ of the southern sky in the \textit{grizY} bands. In addition to this wide-area survey, DES includes a deep supernova survey spanning about 27 deg$^2$, primarily aimed at characterizing type Ia supernova light curves \cite{kessler2015results}.

The main scientific goals of DES are pursued through four complementary probes: type Ia supernovae, large-scale clustering of galaxies, galaxy cluster counts, and weak gravitational lensing \cite{descollaboration2005}. These observations were designed to provide precise photometric redshifts, essential for constraining the equation-of-state parameter $w$ of dark energy.

The survey was conducted over 345 nights across six years. The first data release (DES DR1) included data from the first three years (August 2013 to January 2016), cataloging nearly 400 million astronomical objects \cite{abbott2018dark}. The second data release (DES DR2) incorporates all six years of data, including a reprocessing of DR1 data, and provides a catalog of nearly 700 million objects, with similar proportions of galaxies and stars \cite{abbott2021dark}.

The publicly available DES DR2 dataset was accessed via the DESaccess platform\footnote{\url{https://des.ncsa.illinois.edu/desaccess/}}, an interface that enables users to query and retrieve DES data. We retrieved the data using a Structured Query Language (SQL) query. To select point-like sources, we applied the following constraints,
\begin{lstlisting}[basicstyle=\fontsize{6}{6},lineskip={-1pt}]
WAVG_SPREAD_MODEL_I + 3.0*WAVG_SPREADERR_MODEL_I > 0.005 and
WAVG_SPREAD_MODEL_I + 1.0*WAVG_SPREADERR_MODEL_I > 0.003 and
WAVG_SPREAD_MODEL_I - 1.0*WAVG_SPREADERR_MODEL_I > 0.001 and
WAVG_SPREAD_MODEL_I > -1 and
IMAFLAGS_ISO_I = 0 and
MAG_AUTO_I < 21 ,
\end{lstlisting}
where \lstinline{WAVG_SPREAD_MODEL_I} is a morphological classifier used to distinguish between extended and point-like sources, \lstinline{WAVG_SPREADERR_MODEL_I} represents its associated uncertainty, and \lstinline{IMAFLAGS_ISO_I = 0} ensures that the selected objects are free from artifacts or contamination issues. Additionally, the constraint on \lstinline{MAG_AUTO_I} ensures that only objects with magnitudes \lstinline{I < 21} are included. These selection criteria follow recommendations from the DES DR2 release paper \cite{abbott2021dark} to identify point-like objects, which are expected to have stellar morphology.

\subsection{Sloan Digital Sky Survey}

The Sloan Digital Sky Survey (SDSS) is a spectroscopic and photometric survey that initially operated with a telescope located at the Apache Point Observatory (APO), and since 2017 has also included observations from the Las Campanas Observatory (LCO). The survey has been in operation since 1998, and is currently in its fourth phase, with sixteen data releases. The first data release occurred in 2001, and the most recent one, Data Release 16 (DR16), was published in 2020 \citep{ahumada202016th}. All SDSS data releases are cumulative, meaning that the most recent release contains both the new data and reprocessed data from all previous releases using updated pipelines and improved analyses. Therefore, it is recommended that users access the most recent release to ensure the highest data quality.

The SDSS DR16 dataset provides spectroscopic redshifts and automated spectral classifications—namely, QSO (quasi-stellar object), GALAXY, and STAR—along with a subset of objects flagged for visual inspection of their spectra \citep{bolton2012spectral}. In this work, we utilize these spectral classifications to identify and label the corresponding photometric objects from DES DR2. Through the SkyServer platform, objects detected by optical spectroscopy were retrieved from the SpecObj view of the \texttt{specObjAll} table, which excludes data considered bad or duplicated.

\begin{figure}
    \centering
    \includegraphics[width=0.48\textwidth]{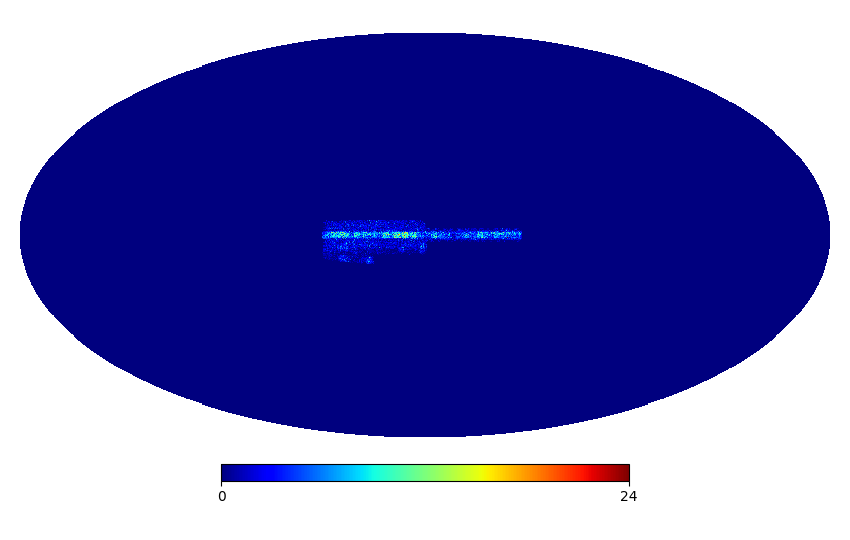}
\caption{Positional cross-match between photometric point-like objects from DES DR2 and spectroscopically classified objects from SDSS DR16. The matched dataset provides the basis for training and testing our classification model and for photometric redshift estimation.}
    \label{fig.data_match_overview}
\end{figure}

\subsection{Positional Cross-Match of DES DR2 Photometry and SDSS DR16 Spectroscopy}

A positional cross-match was performed between the DES DR2 catalog of point-like objects and the SDSS DR16 spectroscopic catalog. Fig. \ref{fig.data_match_overview} provides a visual representation of the resulting matched dataset, highlighting its sky coverage. This procedure utilized the initial DES DR2 catalog, containing 50,679,391 photometric objects, and the full spectroscopic dataset from SDSS DR16, which comprises 5,107,045 spectroscopically detected objects. The cross-match resulted in a final sample of 177,670 matched objects with available spectroscopic classifications and redshifts, forming the basis for further analysis in this work. The composition of this matched catalog, alongside the initial counts from each survey, is summarized in Table \ref{tab.object_counts}. Notice that for the matched sample most of the non-star objects are quasars due to the point-like selection on the photometric catalog. For cosmological purposes, it is not necessary to distinguish between quasars and galaxies; therefore, throughout the text, we refer to quasars as quasars/galaxies with point-like morphology.

The cross-matching process was based on the right ascension $\alpha$ and declination $\delta$ coordinates defined in the equatorial system. We used the Python library \texttt{esutil}\footnote{\url{https://esutil.readthedocs.io/}}, which applies the Hierarchical Triangular Mesh (HTM) indexing method to divide the spherical surface into equal-area triangles, enabling efficient searches for corresponding objects across surveys \cite{kunszt2001hierarchical}. 

A maximum positional tolerance of 1 arcsecond (equivalent to $0.0002778$ degrees) was adopted, allowing at most one match per object. This value is consistent with the average point spread function (PSF) of the DES detection bands, which ranges from 0.83 to 1.1 arcseconds \cite{abbott2021dark}. The PSF reflects the typical positional uncertainty of detected objects; therefore, choosing a tolerance smaller than 1 arcsecond could lead to missed genuine counterparts, while a larger tolerance might increase the number of false matches.

\begin{table}
    \centering
    \resizebox{\linewidth}{!}{
    \begin{tabular}{|l|c|c|c|}
        \hline
        \textbf{Category} & \textbf{DES DR2} & \textbf{SDSS DR16} & \textbf{Matched Sample} \\
        \hline
        Total Objects     & 50,679,391           & 5,107,045              & 168,738              \\
        Stars             & N/A                      & 1,041,130              & 85,896               \\
        Galaxies          & N/A                      & 2,963,274              & 4,649                \\
        Quasars           & N/A                      & 1,102,641              & 78,193               \\
        \hline
    \end{tabular}
    }
    \caption{Summary of object counts across catalogs and in the matched sample.}
    \label{tab.object_counts}
\end{table}

\section{Quasar Candidate Selection}
\label{sec.quasar_selection}

\begin{figure}
    \centering
    \begin{subfigure}{0.48\textwidth}
        \centering
        \includegraphics[width=\textwidth]{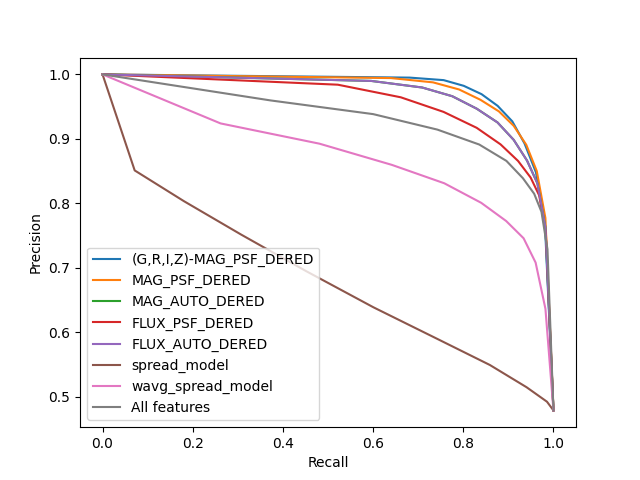}
        \caption{Comparison of feature types}
        \label{fig.pr_curve_1}
    \end{subfigure}
    \hfill
    \begin{subfigure}{0.48\textwidth}
        \centering
        \includegraphics[width=\textwidth]{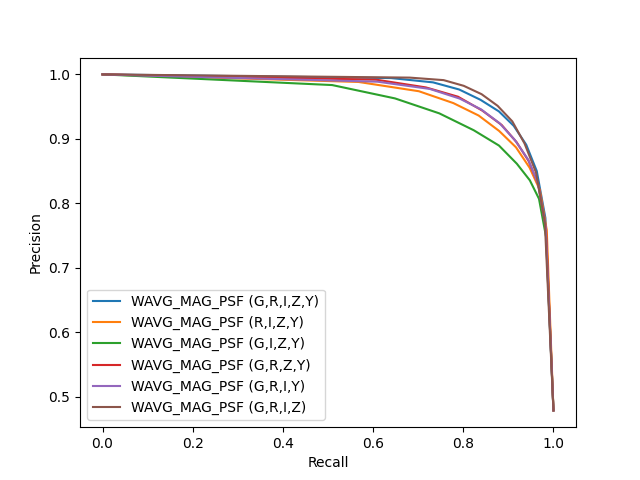}
        \caption{Comparison of magnitude band combination}
        \label{fig.pr_curve_2}
    \end{subfigure}
    \caption{Precision-recall curves for the KNN classification of quasars/galaxies versus stars using different feature combinations, evaluated on the test set. Top: Comparison of feature sets including magnitudes, fluxes, and spread model parameters. The results show that magnitudes outperform other feature types for this classification task. Bottom: Comparison of different combinations of magnitude bands. The PSF magnitudes in the G, R, I, and Z bands provide the best trade-off between precision and recall.}
    \label{fig.precision_recall}
\end{figure}

\begin{figure}
    \centering
    \includegraphics[width=0.48\textwidth]{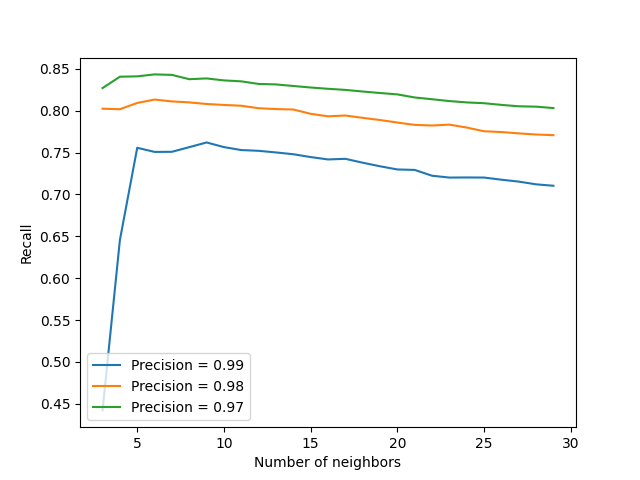}
    \caption{Recall as a function of the number of neighbors $k$ for the K-Nearest Neighbors (KNN) model evaluated on the test set. The curves correspond to different precision thresholds: blue for 0.99, orange for 0.98, and green for 0.97. The highest recall at precision 0.99 is achieved with $k = 11$ neighbors.}
    \label{fig.recall_vs_neighbors}
\end{figure}

\begin{figure}
\centering
\begin{subfigure}{0.45\textwidth}
    \includegraphics[width=\textwidth]{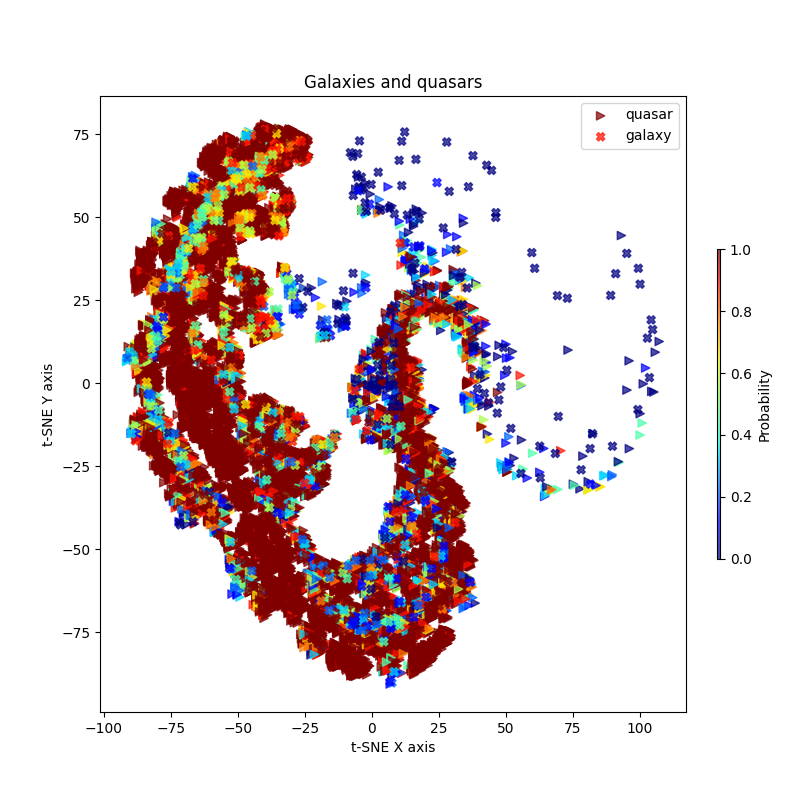}
    \caption{Non-stars (Quasars/Galaxies)} 
    \label{fig.tsne_nonstars}
\end{subfigure}
\hfill
\begin{subfigure}{0.45\textwidth}
    \includegraphics[width=\textwidth]{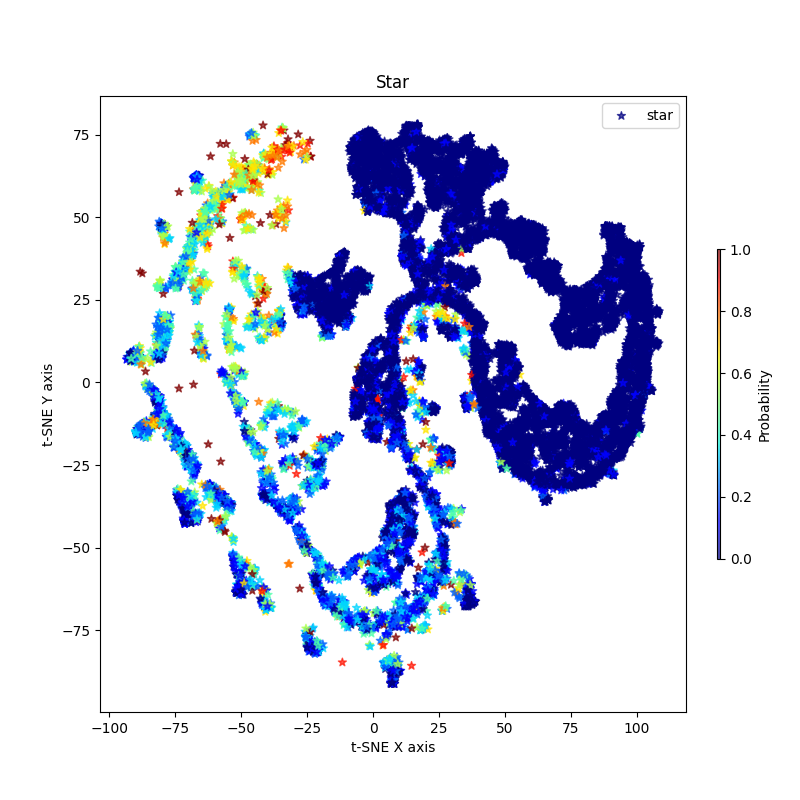}
    \caption{Stars}
    \label{fig.tsne_stars} 
\end{subfigure}
\caption{t-SNE visualization of the feature space used for object classification. The plots show the two-dimensional embedding of (a) photometrically classified non-stars (quasars and galaxies) and (b) photometrically classified stars from the test set, based on their PSF magnitudes in the G, R, I, and Z bands. The distinct clustering illustrates the separability between the two classes achieved by the selected features. The color of each point indicates the probability assigned by the KNN model of being classified as a quasar/galaxy, showing that most objects are confidently classified. }
\label{fig.tsne_visualization} 
\end{figure}

\begin{table}
\centering
\begin{tabular}{lr}
\toprule
\textbf{Dataset} & \textbf{Number of objects}  \\
\midrule
Total      & 168,738  \\
Training Set (75\%) & 126,553  \\
Test Set (25\%)     & 42,185   \\
\bottomrule
\end{tabular}
\caption{Number of objects used for training the classification model of Quasar selection.}
\label{tab.class_split}
\end{table}

We divided the dataset from the cross-match between DES DR2 and SDSS DR16 into two subsets: 75\% for training and 25\% for testing (Table \ref{tab.class_split}). The aim is to train a machine learning model to classify objects as quasars/galaxies or stars. A reliable separation is crucial, as stars contaminate quasar samples used in cosmology. This section describes our approach: the classification model, the evaluation metrics, the selection of features, and the tuning of model parameters.

\subsection{Classification Method: K-Nearest Neighbors}
\label{sec.classifcation_method}

For this classification task, we employed the K‑Nearest Neighbors (KNN) algorithm, a simple yet effective supervised method commonly used in classification problems \citep{cover1967nearest}. The core idea is that objects close together in feature space tend to share the same class.

Given a query object, KNN identifies its $k$ nearest neighbors in the training set—typically using Euclidean distance—and assigns a class label via majority vote. It also provides class probabilities based on the fraction of neighbors in each class. Because KNN is non-parametric and instance-based, it adapts well to complex, non-linear class boundaries like those between stars and quasars/galaxies in photometric feature space.

We used the widely adopted scikit-learn implementation\footnote{\url{https://scikit-learn.org/stable/modules/generated/sklearn.neighbors.KNeighborsClassifier.html}} \cite{pedregosa2011scikit}, which includes efficient scaling and voting routines. KNN has been successfully applied in astronomical classification tasks \cite{ball2007robust, viquar2019machine, li2008k}. Because KNN relies on distance metrics, it is sensitive to feature scaling. To ensure fair weighting across attributes, we normalized each input feature to the range [0,1] using scikit-learn's {\tt MinMaxScaler}\footnote{\url{https://scikit-learn.org/stable/modules/generated/sklearn.preprocessing.MinMaxScaler.html}}. This prevents variables with larger numerical ranges from dominating the distance computation.

\subsection{Metrics}
\label{sec.metrics}

To evaluate the performance of our classification models, we adopted the precision-recall (PR) curve as the main metric. This choice reflects the specific requirements of our study, where it is critical to build a quasar candidate catalog with minimal contamination from stars. In this context, precision and recall provide complementary measures of model performance. 

Precision is defined as the fraction of objects classified as quasars or galaxies that are truly quasars or galaxies. In other words, it quantifies the reliability of positive predictions and is given by
\begin{align}
\text{Precision} = \frac{\text{True Positives}}{\text{True Positives} + \text{False Positives}}.    
\end{align}
A high precision value indicates that the majority of the objects identified as quasars or galaxies are genuine, which is essential for cosmological analyses, as contamination by stars would compromise measurements of large-scale structure and bias scientific conclusions.

Recall, on the other hand, measures the fraction of actual quasars or galaxies that are correctly identified by the classifier. It is defined as
\begin{align}
\text{Recall} = \frac{\text{True Positives}}{\text{True Positives} + \text{False Negatives}}.    
\end{align}
A high recall means that most true quasars and galaxies present in the dataset are recovered by the model, ensuring completeness of the selected sample.

The precision-recall curve illustrates the trade-off between these two quantities as the classification threshold is varied. Our priority is to achieve high precision, so that the final quasar catalog is as free as possible from stellar contaminants. Such contaminants could otherwise introduce significant systematics in cosmological studies.

Although receiver operating characteristic (ROC) curves are commonly used in classification problems, they are less informative in the presence of class imbalance because they give equal weight to both positive and negative classes. In contrast, the precision-recall curve focuses on the performance with respect to the positive class. In what follows, we use the PR curve to guide the choice of model configurations and thresholds that balance precision and recall according to the scientific needs of our study.

\subsection{Feature Selection}
\label{sec.feaure_selection}

To improve the performance of our classification model, we tested different combinations of features from the photometric data to identify which provided the best separation between quasars/galaxies and stars.  Fig.~\ref{fig.pr_curve_1} compares the precision-recall (PR) curves obtained when training the KNN classifier with different sets of features. In this comparison, we considered subsets that included combinations of magnitudes, fluxes, and the spread model parameters. 

Although the spread model is useful for distinguishing point-like and extended sources in general, we found that using it as a feature for training the classifier led to poorer results. This outcome is expected because our original DES dataset was pre-selected to include only objects with point-like morphology by applying a cut on the spread model. As a result, the remaining sample no longer contained the morphological diversity that the spread model would normally help to separate, and the spread model provided little additional information for machine learning classification in this context.

In contrast, we found that the magnitudes of the objects were the most effective features for classification, achieving higher precision at the same level of recall compared to other feature sets. In Fig.~\ref{fig.pr_curve_2}, we focused on comparing different combinations of magnitude bands to determine which subset offered the best performance. We found that using the PSF magnitudes in the G, R, I, and Z bands provided the highest precision-recall curve.

To illustrate the separability of quasars/galaxies and stars within the chosen feature space, defined by the PSF magnitudes in the G, R, I, and Z bands, we applied the t-distributed Stochastic Neighbor Embedding (t-SNE) technique \citep{van2008visualizing} to project the objects in the test set into two dimensions. The resulting visualization reveals distinct clusters for the two classes, highlighting the effectiveness of the selected features for classification. The color of each point represents the probability assigned by the KNN model for the object to be a quasar or galaxy, indicating that most objects are classified with high confidence.

\subsection{Model Tuning}
\label{sec.model_tuning}

Based on the selected features, we first fixed the desired precision at 0.99 to ensure that the final quasar candidate catalog would have minimal contamination from stars. Our goal was to maximize recall while maintaining this high level of precision. To achieve this, we tuned the hyperparameter $k$ (the number of neighbors in the KNN model) and evaluated its impact on the precision-recall performance. As shown in Figure~\ref{fig.recall_vs_neighbors}, the optimal number of neighbors was found to be $k = 11$, which provided the highest recall compatible with the chosen precision. This tuning process also determined the optimal probability threshold, which we found to be 0.8 in order to achieve the target precision. The final model achieved a recall of 0.77 at precision 0.99, with an area under the precision-recall curve (AUC-PR) of 0.975. We then applied this optimized KNN model to classify the full DES DR2 photometric dataset, enabling the identification of new quasar candidates based solely on photometric information.

\section{Photometric Redshift Estimation through Hybrid Machine Learning Approaches}
\label{sec.photo_z}

\begin{figure}
\centering
\begin{subfigure}[t]{0.48\textwidth}
\centering
\includegraphics[width=\textwidth]{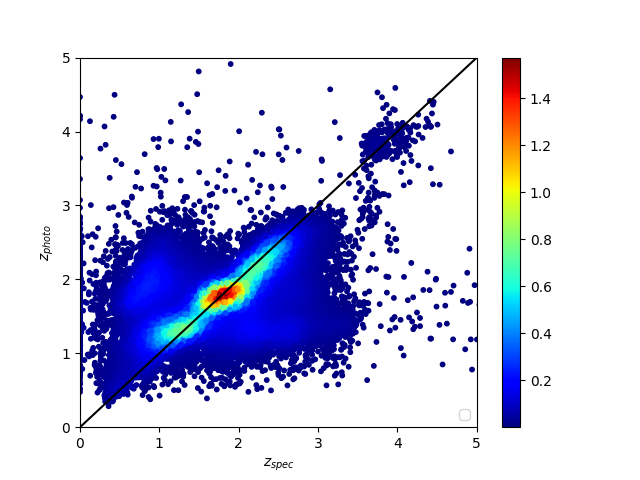}
\caption{Photo-$z$ before outlier removal}
\label{fig.photoz_combined}
\end{subfigure}
\hfill
\begin{subfigure}[t]{0.48\textwidth}
\centering
\includegraphics[width=\textwidth]{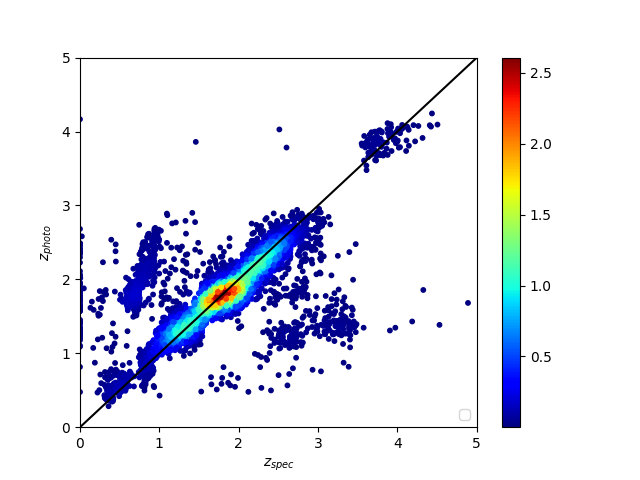}
\caption{Photo-$z$ after outlier removal}
\label{fig_photo_z_outlier_removed}
\end{subfigure}
\caption{Top: photometric redshift estimation results shown for the test set. Top: Photometric redshift estimation combining two machine learning techniques: a Boosted Decision Tree (BDT) from {\tt ANNz} and a Decision Tree Regressor (DTR) from scikit-learn. Note the presence of two clusters off-diagonal which are referred as outliers. Bottom: Photometric redshift after applying a stacked outlier classifier. The fraction of outliers are significantly reduced. }
\label{fig.photoz_hybrid_outlier}
\end{figure}

\begin{figure}
\centering
\includegraphics[width=0.48\textwidth]{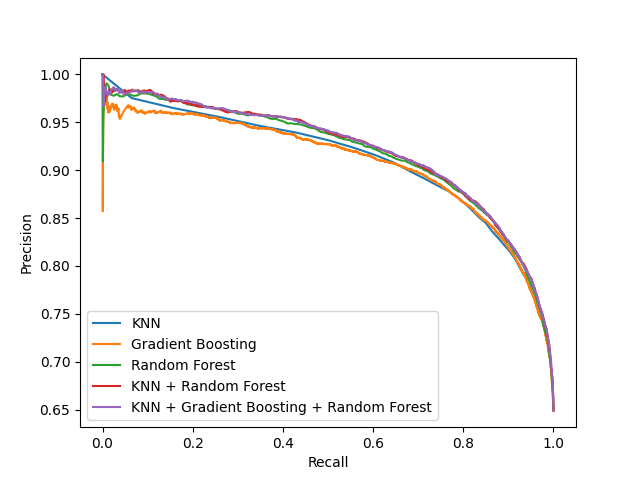}
\caption{Precision-Recall curves for different outlier classifiers evaluated on the test set. The blue, orange, and green curves correspond to individual classifiers: k-Nearest Neighbors (kNN), Gradient Boosting (GB), and Random Forest (RF), respectively. The red curve represents a stacked ensemble combining kNN and RF, while the purple curve shows the performance of the full stacked model combining kNN, GB, and RF. The stacked classifiers, particularly the full ensemble (purple), achieve higher AUC-PR, indicating more effective identification of catastrophic outliers while preserving sample completeness. The full ensemble achieves an AUC-PR of 0.92.} 
\label{fig_outlier_pr_curve}
\end{figure}

Photometric redshift estimation provides redshift information for large samples of galaxies and quasars using multi-band photometry.
While traditional template-fitting methods \citep{benitez2000bayesian, arnouts2011lephare, ilbert2006accurate, bolzonella2000photometric} have been widely used, machine learning offers a powerful alternative by learning non-linear relationships between photometric features and redshift directly from spectroscopic training data \citep{collister2004annz, carliles2010random, cavuoti2015photometric, hoyle2016comparison, Oyaizu_2008, almosallam2016gpz}.
These methods have demonstrated improved accuracy and reduced catastrophic outlier rates in photometric redshift studies, particularly for challenging quasar samples \citep{bovy2012photometric, richards2001photometric, brescia2013photometric}.
The precision of photo-z estimation is crucial for cosmological analyses, as errors can significantly impact weak lensing measurements and large-scale structure constraints \citep{joudaki2018kids, hildebrandt2017kids, prat2018dark}.
In this section, we describe the machine learning models and tools used in our analysis.

\begin{itemize}
\item \textbf{Artificial Neural Networks (ANNs)} are a class of machine learning models inspired by the structure and function of the human brain.
They consist of multiple layers of interconnected nodes, or neurons,'' where each connection has an associated weight. During training, these weights are adjusted to minimize the difference between the network's predictions and the true values. ANNs are highly capable of learning complex, non-linear relationships between input features (like photometric magnitudes and colors) and output targets (such as redshift). Their ability to model intricate patterns makes them particularly effective for photometric redshift estimation, as demonstrated in surveys like 2dFGRS and SDSS \citep{collister2004annz, cavuoti2015photometric, newman2013deep2}. 
\item \textbf{Decision Trees (DTs)} are non-parametric supervised learning models used for both classification and regression. They operate by recursively partitioning the feature space into a set of rectangular regions. Each internal node in the tree represents a decision'' based on a threshold applied to a single feature, and each branch corresponds to the outcome of that decision.
This process continues until a leaf node is reached, which represents the predicted output (e.g., a redshift value).
Decision Trees offer a high degree of interpretability due to their clear, rule-based structure.
However, a single decision tree can be prone to overfitting, especially when dealing with noisy data or complex relationships, if not properly regularized or constrained \citep{carliles2010random}.
\item \textbf{Boosted Decision Trees (BDTs)} are an ensemble learning method that combines the predictions of multiple weak'' decision tree learners to create a single, strong predictive model. Unlike Random Forests, BDTs build trees sequentially. Each new tree is constructed to focus specifically on the instances that were mispredicted or had large errors by the ensemble of all previously built trees. This iterative process of boosting'' allows the model to progressively correct its errors, leading to improved accuracy and reduced bias compared to a single decision tree.
BDTs have demonstrated consistently strong performance in various photometric redshift estimation tasks in large astronomical surveys, including SDSS and DES \citep{gerdes2010boosted, hoyle2016comparison}.
\item \textbf{Random Forest Regressors (RFRs)} are another powerful ensemble learning method that builds upon the concept of decision trees.
A Random Forest constructs multiple decision trees during training, but with two key differences from basic Decision Trees: (1) each tree is built using a bootstrap sample (a random sample with replacement) of the training data, and (2) at each split in a tree, only a random subset of features is considered.
These two sources of randomness help to decorrelate the individual trees within the forest.
The final prediction is then obtained by averaging the predictions of all individual trees.
This aggregation process significantly reduces overfitting and decreases the variance of the model's predictions.
Random Forests have been widely adopted for photometric redshift estimation, offering robust performance across heterogeneous datasets and providing good generalization capabilities \citep{carliles2010random}.
\item \textbf{Gradient Boosted Regressors (GBRs)} are an extension of Boosted Decision Trees that utilize gradient descent optimization.
Similar to BDTs, GBRs build an ensemble of decision trees sequentially.
However, instead of simply correcting errors from previous trees, each new tree in a GBR is trained to predict the residuals'' or errors'' of the ensemble's current prediction with respect to the true target values.
This process is framed as minimizing a specified loss function (e.g., mean squared error for regression) through gradient descent at each iteration.
By iteratively reducing the errors in this gradient-based manner, GBRs are capable of yielding highly accurate and robust models that are particularly well-suited for complex photometric redshift relationships and large, noisy datasets \citep{hoyle2016comparison}.
\item \textbf{K-Nearest Neighbors Regressors (KNRs)} are a simple, non-parametric, instance-based learning algorithm.
They are the regressor version of the classifier described in Sec.~\ref{sec.classifcation_method}.
To predict the redshift of a new, unseen object, KNRs identify the $k$ closest data points (neighbors) to that object in the multi-dimensional feature space (defined by photometric parameters).
The prediction for the new object's redshift is then typically calculated as the average of the redshifts of its $k$ nearest neighbors.
KNRs are conceptually straightforward and require no explicit training phase (they are ``lazy learners'' as computation is deferred until prediction time).
While they can be computationally intensive for very large datasets due to the need to calculate distances to all training points, KNRs often serve as effective baselines in photometric redshift studies, providing a simple yet robust approach to inferring redshifts based on local similarity in the photometric space \citep{ball2007robust}.
\end{itemize}

\subsection{Implementation and Performance}
\label{sec.photoz_tools}

\begin{table*}
\centering
\caption{Training and test sets used for photometric redshift regression models and outlier removal classification models.}
\label{tab.photoz_split}
\begin{tabular}{lr}
\toprule
\textbf{Dataset} & \textbf{Number of objects} \\
\midrule
Total Quasar/Galaxy sample (100\%) & 62,570 \\
Photo-z Training Set (40\%) & 25,028 \\
Photo-z Test Set (60\%) & 37,542 \\
\midrule
Outlier classifier Training Set (50\% of Test set) & 18,771 \\
Outlier classifier Test Set (50\% of Test Set) & 18,771 \\
\bottomrule
\end{tabular}
\end{table*}

To estimate photometric redshifts (photo-$z$) for quasars, we applied several machine learning regression models. These models were trained on a subset of the matched sample consisting of objects classified as quasars or galaxies by our classification model. The dataset was then split into 40\% for training and 60\% for testing. We allocated a larger portion to the test set because it was further subdivided for training and testing the outlier removal classifier (see Table~\ref{tab.photoz_split}). We focused on two computational frameworks for the redshift estimation: 
\begin{itemize}
    \item \textbf{{\tt ANNz}} \citep{collister2004annz, lahav2012annz} is a tool designed for photometric redshift estimation in astronomy. It supports Artificial Neural Networks and Boosted Decision Trees, and provides functionalities for probability distribution estimation and flexible input feature selection. Both ANNs and BDTs were trained using {\tt ANNz} with photometric magnitudes and, in a separate configuration, color indices derived from these magnitudes.

    \item \textbf{Scikit-learn} \citep{pedregosa2011scikit} is a widely-used Python library for machine learning. We used scikit-learn to implement Decision Tree Regressors, Gradient Boosted Regressors, Random Forest Regressors, and K-Nearest Neighbors Regressors, all trained on photometric magnitudes.
\end{itemize}

Fig. \ref{fig.photoz_combined} compares the photometric redshift with the spectroscopic one for the test set. Across the best models, we observe a prominent main cluster of objects spanning approximately $0.5 < z < 3$, where the photometric redshift estimates show small errors and closely track the spectroscopic redshifts. Additionally, two distinct off-diagonal clusters are present; these correspond to objects for which the photometric redshift is not accurately recovered and are thus classified as outliers. Notably, we also identify a cluster near $z \approx 4$, where photometric redshifts are well recovered. The outliers do not extend beyond $z \approx 3.5$, meaning the high-redshift population around $z \approx 4$ remains reliably identified.

We compared different ML models implemented in both {\tt ANNz} and {\tt scikit-learn} frameworks. The feature space chosen for photometric redshift estimation includes four color indices—\lstinline{MAG_AUTO_G-R_DERED}, \lstinline{MAG_AUTO_R-I_DERED}, \lstinline{MAG_AUTO_I-Z_DERED}, and \lstinline{MAG_AUTO_Z-Y_DERED}—along with the dereddened r-band magnitude \lstinline{MAG_AUTO_R_DERED} and a morphological parameter, \lstinline{KRON_RADIUS}. In \lstinline{ANNz}, the best-performing model was the BDT, while in scikit-learn, the DT provided superior performance over GBR and RFR. This preference arises because while models like GBR and RFR aim to reduce overall prediction errors, they do not effectively distinguish between outliers and well-predicted objects. For our purposes, achieving a large group of non-outliers is more valuable than minimizing individual prediction errors uniformly. Further discussion on the photo-$z$ estimation with different ML models is provided in Appendix \ref{sec.comparison of photo_z methods}.

The best overall performance (Fig. \ref{fig.photoz_combined}) was obtained with the hybrid model, which combines the {\tt ANNz} optimized BDT and the scikit-learn DTR using a $k$-nearest neighbors (kNN)-based dynamic weighting scheme. This hybrid approach leverages the complementary strengths of both models, resulting in a large main cluster with small photometric redshift errors, reduced size of outlier clusters, and preservation of the high-redshift $z \approx 4$ cluster.

\subsection{Outlier Removal}
\label{sec.outlier_removal}

Photometric redshift estimation is particularly susceptible to catastrophic outliers—objects where the difference between the predicted and true redshift is large—often due to degeneracies in color–redshift space or the impact of photometric noise \citep{richards2002spectroscopic, ball2007robust, bovy2012photometric}. We observed that groups of catastrophic outliers are predominantly concentrated at intermediate redshifts up to $z \approx 3.5$ with a noticeable reduction in outlier frequency at higher redshifts. This trend is consistent with \citep{wu2004color, han2016photometric, yang2017quasar, peters2015quasar}.

To address this, we implemented a secondary step aimed at identifying and filtering such outliers. We defined catastrophic outliers as objects satisfying 
\begin{align}
\vert z_{\rm photo} - z_{\rm spec} \vert \ge 0.4 \,.    
\end{align}
To train and validate our classifier, we reserved 50\% of the original test set for this task.

We compared several classification algorithms and found that the best performance was achieved by a stacked ensemble classifier\footnote{\url{https://scikit-learn.org/stable/modules/generated/sklearn.ensemble.StackingClassifier.html.}} A stacked ensemble combines the outputs of multiple base classifiers to produce a final prediction, leveraging the strengths of each individual model to improve overall performance \citep{wolpert1992stacked}. Our ensemble integrated k-Nearest Neighbors (kNN), Gradient Boosting Machine (GBM), and Random Forest (RF) classifiers. The base models—kNN, GBM, and RF—were previously employed in our work as regressors to estimate photometric redshifts. In this classification setting, however, these models are repurposed to assign an outlier probability to each object, which is then used to distinguish likely outliers from the main quasar population

To assess performance, we used precision–recall (PR) curves for all classifiers, including both individual models and stacked ensembles. The PR curves, shown in Figure~\ref{fig_outlier_pr_curve}, highlight how each classifier balances precision and recall, with the stacked models—especially the full ensemble—achieving the best trade-off, with an Area Under the Curve (AUC) of 0.92 We fixed the threshold of the classification at 0.79 yielding a precision of 0.9 and recall of 0.73. After applying the outlier classification to the full test sample, the filtered dataset (Figure~\ref{fig_photo_z_outlier_removed}) exhibited a significantly reduced fraction of catastrophic outliers compared to the original predictions, demonstrating the effectiveness of the outlier rejection step.

\subsection{Evaluation of photometric error}
\label{sec.error_photo_z}

\begin{figure}
    \centering
    \includegraphics[width=\linewidth]{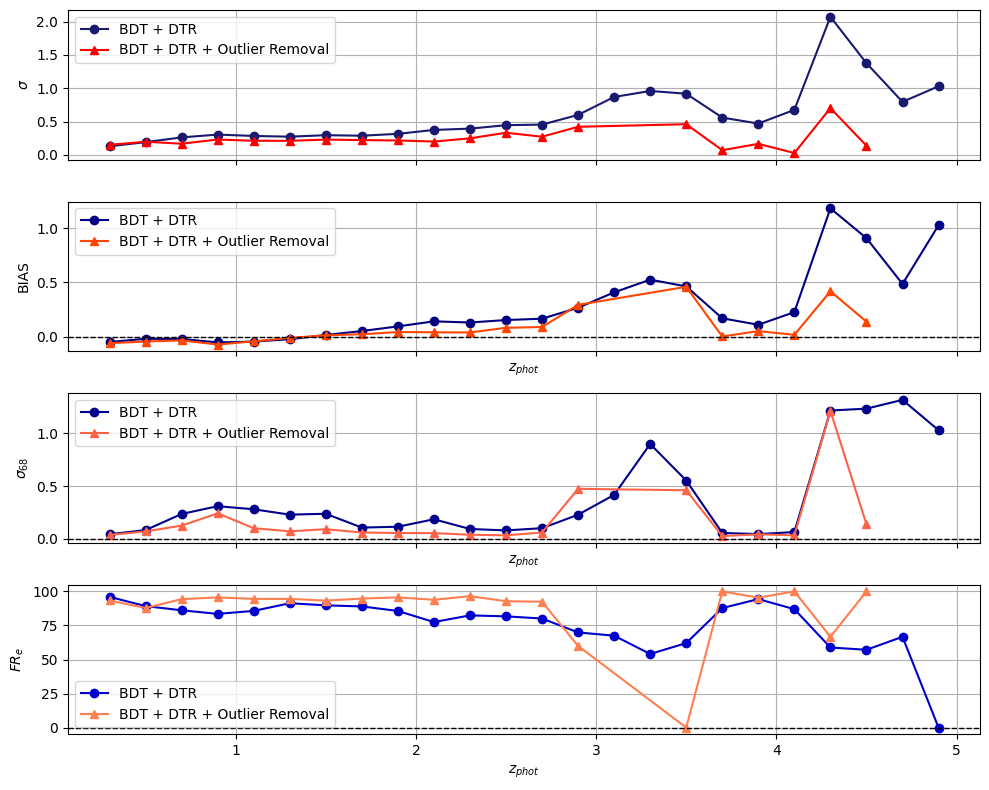}
    \caption{Evaluation of photometric redshift error metrics as a function of spectroscopic redshift $ z_{\mathrm{spec}} $. From top to bottom: (1) scatter $ \sigma $, (2) bias, (3) 68th percentile error $ \sigma_{68} $, and (4) effective retention fraction $ FR_e $. Two models are shown: a baseline BDT + DTR hybrid model (blue circles) and the same model with an additional outlier removal step (red triangles). The bias remains close to zero up to $ z \sim 1.5 $, where the training data are most concentrated. All error metrics degrade for $ z \gtrsim 4.5 $ due to sample sparsity. The outlier removal procedure leads to a consistent reduction in scatter and error metrics across most of the redshift range. The retention fraction $ FR_e $ is generally higher for the model with outlier removal, except near $ z \sim 3.5 $, where limited data causes fluctuations in performance. }
    \label{fig.photo_z}
\end{figure}

To quantify the performance of our photometric redshift estimator, we computed four metrics as functions of the estimated redshift $ z_{\mathrm{phot}} $: the bias, the scatter $ \sigma $, the 68th percentile error $ \sigma_{68} $, and the effective retention fraction $ FR_e $. These metrics are defined respectively as
\begin{align}
\text{Bias} &= \left\langle \frac{z_{\mathrm{phot}} - z_{\mathrm{spec}}}{1 + z_{\mathrm{spec}}} \right\rangle\quad , \\
\sigma &= \sqrt{ \left\langle \left( \frac{z_{\mathrm{phot}} - z_{\mathrm{spec}}}{1 + z_{\mathrm{spec}}} \right)^2 \right\rangle }\quad , \\
\sigma_{68} &= \text{68th percentile of } \left| \frac{z_{\mathrm{phot}} - z_{\mathrm{spec}}}{1 + z_{\mathrm{spec}}} \right| \quad ,  \\
FR_e &= \frac{100}{n} \sum_{i=1}^{n} \delta_i \quad , \quad \delta_i = 
\begin{cases}
1 & \text{if } \left| \frac{z^{i}_{\mathrm{phot}} - z^{i}_{\mathrm{spec}}}{1 + z^{i}_{\mathrm{spec}}} \right| < \epsilon \\
0 & \text{otherwise} 
\end{cases} \,.
\end{align}

Here, $ n $ is the total number of objects in the redshift bin. The \textit{bias} measures the average systematic offset between photometric and spectroscopic redshifts, indicating whether predictions tend to over- or underestimate the true values. The \textit{scatter} $ \sigma $ captures the overall spread in the normalized residuals, while $ \sigma_{68} $ estimates the width of the core distribution. The \textit{effective retention fraction} $ FR_e $ reflects the percentage of objects retained within the outlier threshold $ \epsilon = 0.15 $, thus quantifying robustness to catastrophic errors.

Figure~\ref{fig.photo_z} displays the evolution of these metrics across the redshift range for two models: the baseline BDT + DTR ensemble, and the same model with a post-processing outlier removal step. In the top three panels, we observe that the outlier removal consistently reduces the scatter $ \sigma $, bias, and $ \sigma_{68} $ across nearly the entire redshift range. The improvements are especially notable at higher redshifts ($ z \gtrsim 2 $), where the baseline model exhibits increasing deviation and wider error distributions. The outlier removal smooths these fluctuations and stabilizes performance, indicating improved resilience in low-data regimes.

The bias remains close to zero up to $ z \sim 1.5 $ for both models, where the bulk of the training set is concentrated, suggesting effective calibration. Beyond this range, the baseline model begins to exhibit increasing bias, whereas the model with outlier removal maintains near-zero bias up to $ z \sim 3 $, diverging only slightly thereafter. Around $ z \sim 3.5 $, all error metrics peak, with the baseline model performing significantly worse. Interestingly, at $ z \sim 4 $, the metrics dip again, particularly for the model with outlier removal, which suggests it better handles the absence of clustered outliers in that region.

The bottom panel shows the effective retention fraction $ FR_e $. The model with outlier removal maintains a consistently high retention fraction above 80\% across most of the redshift range, slightly outperforming the baseline except near $ z \sim 3.5 $, where both models exhibit a temporary drop likely due to data sparsity. The retention peak at $ z \sim 4 $ aligns with the dip in error metrics, again indicating that outlier rejection enhances both precision and robustness without excessively discarding valid predictions.

\section{Building the DES photometric quasar catalog}
\label{sec.applying_on_des}

With the full machine learning pipeline trained and validated, we proceeded to apply it to the Dark Energy Survey (DES) DR2 photometric data to generate a new, large-scale catalog of quasars. The process followed a multi-stage approach to ensure the quality and reliability of the final sample.

First, we began with the full catalog of approximately 50 million point-like sources. From this initial set, we applied a selection cut based on magnitudes and morphological parameters to isolate objects within a well-populated and reliable region of feature space—where the training set offers robust coverage and our models are expected to perform optimally. Specifically, we restricted the data to lie within the 1st to 99th percentile range of the training set distributions for each feature. This minimizes the impact of outliers and ensures consistency between the training and prediction domains. The applied cuts were:
\begin{align}
    18.55 &> \tt WAVG\_MAG\_PSF\_G\_DERED > 22.75 \,, \\
    18.43 &> \tt WAVG\_MAG\_PSF\_R\_DERED > 22.35 \,, \\
    18.38 &> \tt WAVG\_MAG\_PSF\_I\_DERED > 22.13 \,, \\
    18.27 &> \tt WAVG\_MAG\_PSF\_Z\_DERED > 21.96 \,, \\
    18.25 &> \tt WAVG\_MAG\_PSF\_Y\_DERED > 21.70 \,, \\
    3.5 &> \tt KRON\_RADIUS > 3.88 \,.
\end{align}

In order to avoid contamination from bright local structures, we applied a mask to regions of the sky that contain prominent nearby galaxies and dwarf systems. For each source we defined a circular mask centered on the equatorial coordinates given in Table~\ref{tab:sources}, with a radius given by the reported apparent size to ensure the removal of the entire brightness region.  

\begin{table}
\centering
\footnotesize
\setlength{\tabcolsep}{4pt} % tighten column spacing
\caption{Angular sizes and equatorial coordinates of selected sources.}
\label{tab:sources}
\begin{tabular}{lccc}
\hline
Source & Apparent size & RA [$^{\circ}$] & DEC [$^{\circ}$] \\
\hline
NGC300                  & 20.89 arcmin  & 13.72  & $-37.68$ \\
Sculptor Dwarf Galaxy   & 63.10 arcmin  & 15.04  & $-33.71$ \\
Fornax Dwarf Galaxy     & 56.23 arcmin  & 39.99  & $-34.45$ \\
IC1613 (Cetus)          & 15.14 arcmin  & 16.23  & $+2.13$  \\
NGC1399 (Fornax cluster)& 7.59 arcmin   & 54.62  & $-35.45$ \\
Nubecula Major          & 10.75$^{\circ}$ & 80.89  & $-69.76$ \\
\hline
\end{tabular}
\end{table}

Then, we deploy the optimized KNN classifier on this pre-selected sample to identify quasar candidates. This step yielded a catalog of 872,373 potential quasars. Finally, our hybrid photo-z model (Section \ref{sec.photo_z}) was used to estimate photometric redshifts for all of these candidates. Table~\ref{tab.final_catalog_summary} summarizes the number of objects at each stage of the photometric selection process for constructing the quasar catalog.

\begin{table}
\centering
\caption{Summary of the number of objects in the photometric catalog}
\label{tab.final_catalog_summary}
\resizebox{\linewidth}{!}{
\begin{tabular}{lr}
\toprule
\textbf{Sample Description} & \textbf{Number of objects} \\
\midrule
DES point-like sources & 50,679,391 \\
DES point-like sources after color/magnitude cut & 14,986,431 \\
Quasar/Galaxy candidates  & 872,373 \\
Quasar/Galaxy candidates after outlier removal & 675,683 \\
\bottomrule
\end{tabular}
}
\end{table}

\begin{figure}
\centering
\includegraphics[width=0.48\textwidth]{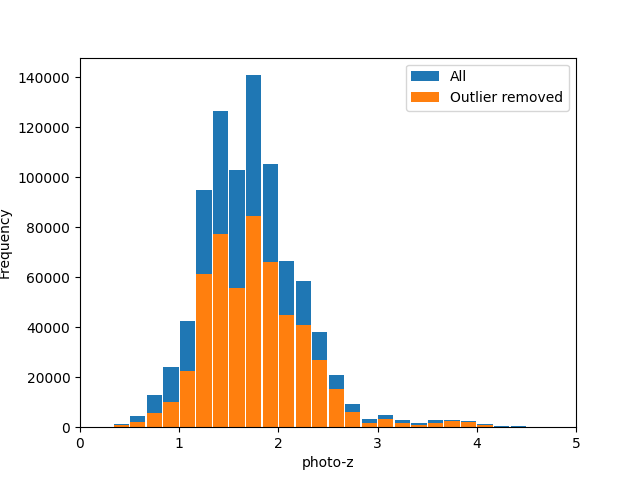}
\caption{Photometric redshift distribution of the DES quasar catalog selected in this work. The photometric redshift were computed with our model that combines the BDT from {\tt ANNz} and a DTR from scikit-learn. The distribution spans a broad redshift range from $z \approx 0.5$ to beyond $z = 3$, with a pronounced peak around $z \sim 1.5$. The blue curve shows the distribution before outlier removal (872,373 objects), while the orange curve shows the distribution after removing catastrophic outliers (675,683 objects). }
\label{fig.des_redshift_dist}
\end{figure}

\begin{figure}
\centering
\includegraphics[width=0.48\textwidth]{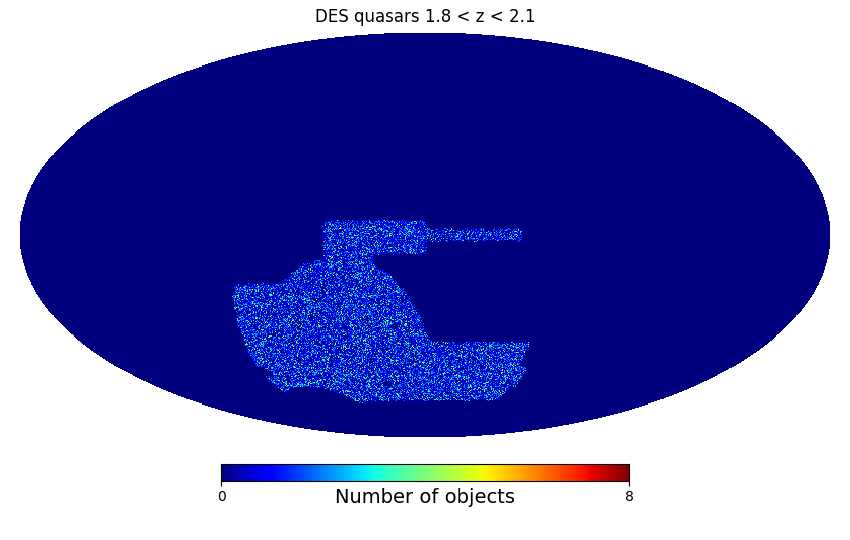}
\caption{Tomographic projection of DES quasars in redshift slice $1.8<z<2.1$. Color scale denotes relative overdensity. The quasars appear to trace the large-scale structure of the Universe, providing maps suitable for cosmological studies through statistical analysis.}
\label{fig.des_tomographic_map}
\end{figure}

The photometric redshift distribution of the final catalog is presented in Fig. \ref{fig.des_redshift_dist}. The distribution spans a wide redshift range from $z \approx 0.1$ to $z > 3$, with a prominent peak at $z \sim 1.5$. The distribution that includes outliers shows a small secondary peak around $z \approx 0$, which corresponds to catastrophic outliers. This issue is corrected in the distribution after outlier removal, where the low-redshift peak disappears.  Fig. \ref{fig.des_tomographic_map} illustrate a tomographic map of the selected quasars. The quasars appear to trace the large-scale structure of the Universe, providing maps suitable for cosmological studies through statistical analysis.

\subsection{High-Redshift Quasar Sample ($z \gtrsim 3.5$)}
\label{sec.high_z}

\begin{figure}
\centering
\includegraphics[width=0.48\textwidth]{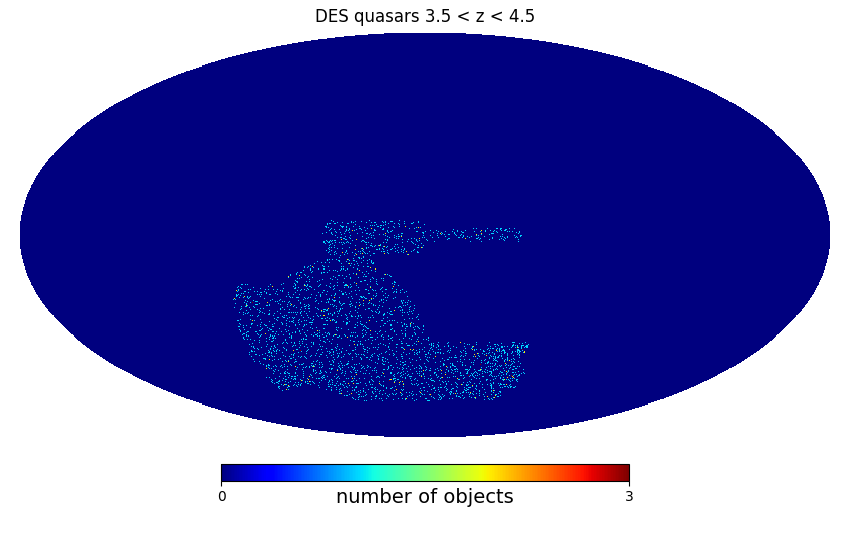}
\caption{Sky distribution of DES quasar candidates with $z \gtrsim 3.5$ (10,756 objects). Despite their low surface density and shot-noise limitation, this high-redshift sample extends the catalog to earlier cosmic times and remains cosmologically valuable.}
\label{fig.des_tomographic_map_high_z}
\end{figure}

\begin{figure}
\centering
\includegraphics[width=0.48\textwidth]{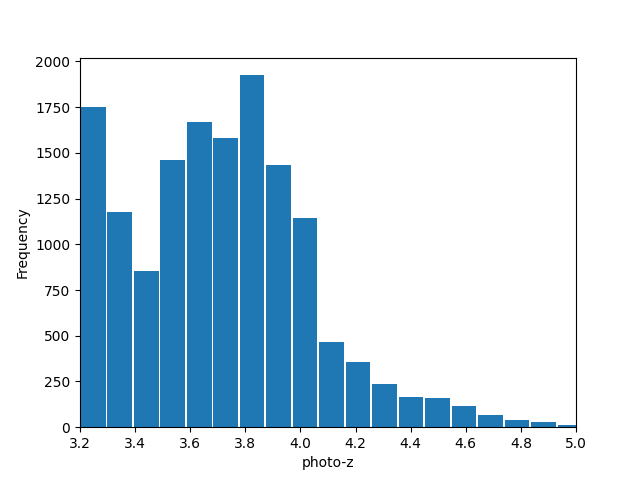}
\caption{Photometric redshift distribution of the DES quasar catalog selected in this work for the high-redshift range $3.3 < z < 5$. This panel highlights the small secondary peak at high redshift. Only the histogram of the complete sample is shown, without outlier removal, due to the low outlier fraction at this range.}
\label{fig.des_redshift_dist_high_z}
\end{figure}

The subset of quasars at redshift $z \gtrsim 4$ represents an important, though comparatively small, part of the DES photometric catalog. While the number of objects identified in this regime is only $10,756$ (for $z \geq 3.5$), these sources remain cosmologically valuable. Quasars at such early times trace the Universe during its first $\sim$1.5 Gyr, providing unique information on the growth of large-scale structure and the transition periods of cosmic reionization. 

A distinctive feature of this regime is that the high-$z$ sample does not require the additional cleaning procedures applied at lower redshifts. This is because the photometric redshift estimates at $z \approx 4$ are robust and do not produce catastrophic outliers beyond $z \approx 3.5$. Thus, the larger catalog can safely be used for high-$z$ studies without risking degradation from misestimated redshifts.

Figure \ref{fig.des_tomographic_map_high_z} shows the sky distribution of quasars with $z \gtrsim 3.5$. Despite the lower density and, therefore, higher shot-noise, this subsample constitutes a reliable and valuable component of the catalog for cosmological applications.

\section{Comparison Between the point-like and extended sources catalogs}

This work complements the study by \cite{Abdalla2025machine}, which performed photometric redshift estimation on galaxy samples from the Dark Energy Survey (DES DR2). The catalog by \cite{Abdalla2025machine} is based on galaxies cross-matched with the VIPERS spectroscopic survey, which is limited to intermediate redshifts, roughly spanning $ 0.5 \lesssim z \lesssim 1.5 $, with a median redshift around $ z \sim 0.7 $. Consequently, their photometric redshift models are optimized for galaxy populations in this regime and are less applicable to higher redshifts.

In contrast, our quasar-focused analysis uses SDSS DR16 as the spectroscopic training set, covering a much broader redshift range from $ z \sim 0 $ up to $ z \sim 6 $. This allows us to probe both low- and high-redshift populations. Notably, our model captures a clean population of high-redshift quasars around $ z \sim 4 $, a region that is beyond the reach of the galaxy-based model from \cite{Abdalla2025machine}.

The performance of photometric redshift estimation also differs between the two works. In \cite{Abdalla2025machine}, the galaxy photo-$ z $ errors are tightly constrained within the redshift interval $ z \lesssim 1.2 $, achieving a 68th percentile error of $ \sigma_{68} \sim 0.035 $ and a catastrophic outlier rate of approximately 3\%. These values reflect the dense and relatively homogeneous nature of the galaxy training set in this range, and the fact that the model is restricted to extended sources with well-defined color–redshift relations.

In our quasar analysis, photo-$ z $ performance is more heterogeneous due to the broader redshift range and the complex spectral energy distributions of quasars, which lead to degeneracies in color space. Nonetheless, we achieve robust predictions with bias and scatter metrics below 0.5 for most of the range up to $ z \sim 3 $, with a notable improvement around $ z \sim 4 $ where model residuals are minimized and the outlier rate drops. Beyond $ z \sim 4.5 $, performance degrades due to the sparsity of training data, as reflected in increased $ \sigma $, $ \sigma_{68} $, and bias, as well as a drop in the effective retention fraction.

These differences highlight the complementary nature of the two catalogs. \cite{Abdalla2025machine} provides a well-calibrated, low-error photo-$ z $ catalog for galaxies in the intermediate-redshift regime, ideal for large-scale structure studies and tomographic analyses. Our quasar catalog, in turn, extends the utility of DES DR2 to higher redshifts and to point-like sources, opening possibilities for cosmological studies involving quasars.

\section{Conclusions}

This paper successfully constructed a robust catalog of point-like objects by performing a positional cross-match between the Dark Energy Survey Data Release 2 (DES DR2) and the Sloan Digital Sky Survey Data Release 16 (SDSS DR16). This initial cross-match resulted in a sample of 168,738 matched objects with available spectroscopic classifications and redshifts. The primary goal was to identify quasars, crucial tracers for cosmological studies, while effectively separating them from contaminating stars.

To achieve this, we employed a K-Nearest Neighbors (KNN) classification model, optimizing its performance by carefully selecting photometric features, specifically PSF magnitudes in the G, R, I, and Z bands. Through hyperparameter tuning, we achieved a recall of 0.77 at a precision of 0.99, ensuring that our final quasar candidate catalog is highly reliable with minimal stellar contamination. 

Beyond accurate classification, this catalog holds significant cosmological potential due to the subsequent photometric redshift (photo-$z$) estimation for the quasar/galaxy objects identified. We utilized a hybrid machine learning approach, combining a color-optimized Boosted Decision Tree (BDT) from {\tt ANNz} with a Decision Tree Regressor (DTR) from {\tt scikit-learn}. This approach allowed us to leverage the strengths of different models, resulting in a catalog with generally small photometric redshift errors. 

A key aspect of our work was addressing the presence of catastrophic outliers in the photometric redshift estimates. We developed and applied a stacked outlier classifier, significantly reducing the fraction of these outliers. This is critical for cosmological analyses, as outliers can introduce substantial biases in measurements of large-scale structure and cosmic evolution.

We highlight the existence of two versions of the catalog: one with a larger fraction of outliers in terms of redshift, and another with a significantly reduced outlier fraction. The larger catalog contains 872,373 objects while the one with reduced outliers contains 675,683 objects. The redshift distribution of the catalog primarily spans from $z \approx 0.5$ to $z > 3$, peaking at $z \sim 1.5$. While the catalog with fewer outliers offers greater precision for most redshift ranges, our analysis revealed that for objects at $z \approx 4$, it is safe to utilize the larger catalog (with 872,373 objects). This is because the high-redshift population around $z \approx 4$ remains reliably identified, with photometric redshifts well recovered and outliers not extending beyond $z \approx 3.5$. However, the number of objects with $z \geq 3.5$ is relatively small, totaling 10,756. The cleaner, reduced-outlier catalog contains 675,683 objects and can safely be used to probe the redshift range $0.5<z<3$. 

This catalog, with its carefully classified and redshift-estimated quasar candidates, provides a valuable resource for future cosmological investigations, especially for probing the late universe. As it complements the galaxy-based catalog presented in \citet{Abdalla2025machine}, future work will focus on combining both datasets to perform cosmological analyses.

\begin{acknowledgements}
P.M.: This study was financed in part by the São Paulo Research Foundation (FAPESP) through grant 2021/08846-2 and 2023/07728-1. Also, this study was financed in part by the Coordenação de Aperfeiçoamento de Pessoal de Nível Superior – Brasil (CAPES) – Finance Code 001. F.B.A thanks the University of Science and Technology of China and the Chinese Academy of Science for grant number KY2030000215. E.A. is supported by a CNPq grant. The authors acknowledge the National Laboratory for Scientific Computing (LNCC/MCTI, Brazil) for providing HPC resources of the SDumont supercomputer, which have contributed to the research results reported within this paper. URL: http://sdumont.lncc.br. 
\end{acknowledgements}

%\section*{Data availability}
%The data underlying this article can be shared upon reasonable request from the corresponding author. They will be freely accessible after the publication of all projects planned by our group from the repository \href{https://github.com/zxcorr/DES_QSO}{\texttt{https://github.com/zxcorr/DES\_QSO}}.

\bibliographystyle{aa}
\bibliography{references_1}

@article{collister2004annz,
  title={ANNz: estimating photometric redshifts using artificial neural networks},
  author={Collister, Adrian A and Lahav, Ofer},
  journal={Publications of the Astronomical Society of the Pacific},
  volume={116},
  number={818},
  pages={345},
  year={2004},
  publisher={IOP Publishing}
}

@article{abdalla2022bingo,
  title={The BINGO project-I. Baryon acoustic oscillations from integrated neutral gas observations},
  author={Abdalla, Elcio and Ferreira, Elisa GM and Landim, Ricardo G and Costa, Andre A and Fornazier, Karin SF and Abdalla, Filipe B and Barosi, Luciano and Brito, Francisco A and Queiroz, Amilcar R and Villela, Thyrso and others},
  journal={Astronomy \& Astrophysics},
  volume={664},
  pages={A14},
  year={2022},
  publisher={EDP Sciences}
}

@article{abbott2019dark,
  title={Dark Energy Survey year 1 results: Joint analysis of galaxy clustering, galaxy lensing, and CMB lensing two-point functions},
  author={Abbott, TMC and Abdalla, FB and Alarcon, A and Allam, S and Annis, J and Avila, S and Aylor, K and Banerji, M and Banik, N and Baxter, EJ and others},
  journal={Physical Review D},
  volume={100},
  number={2},
  pages={023541},
  year={2019},
  publisher={APS}
}

@article{wuensche2022bingo,
  title={The BINGO project-II. Instrument description},
  author={Wuensche, Carlos A and Villela, Thyrso and Abdalla, Elcio and Liccardo, Vincenzo and Vieira, Frederico and Browne, Ian and Peel, Michael W and Radcliffe, Christopher and Abdalla, Filipe B and Marins, Alessandro and others},
  journal={Astronomy \& Astrophysics},
  volume={664},
  pages={A15},
  year={2022},
  publisher={EDP Sciences}
}

@article{abdalla2022bingo-iii,
  title={The BINGO Project-III. Optical design and optimization of the focal plane},
  author={Abdalla, Filipe B and Marins, Alessandro and Motta, Pablo and Abdalla, Elcio and Ribeiro, Rafael M and Wuensche, Carlos A and Delabrouille, Jacques and Fornazier, Karin SF and Liccardo, Vincenzo and Maffei, Bruno and others},
  journal={Astronomy \& Astrophysics},
  volume={664},
  pages={A16},
  year={2022},
  publisher={EDP Sciences}
}

@article{liccardo2022bingo,
  title={The BINGO project-IV. Simulations for mission performance assessment and preliminary component separation steps},
  author={Liccardo, Vincenzo and de Mericia, Eduardo J and Wuensche, Carlos A and Abdalla, Elcio and Abdalla, Filipe B and Barosi, Luciano and Brito, Francisco A and Queiroz, Amilcar and Villela, Thyrso and Peel, Michael W and others},
  journal={Astronomy \& Astrophysics},
  volume={664},
  pages={A17},
  year={2022},
  publisher={EDP Sciences}
}

@article{fornazier2022bingo,
  title={The BINGO project-V. Further steps in component separation and bispectrum analysis},
  author={Fornazier, Karin SF and Abdalla, Filipe B and Remazeilles, Mathieu and Vieira, Jordany and Marins, Alessandro and Abdalla, Elcio and Santos, Larissa and Delabrouille, Jacques and Mericia, Eduardo and Landim, Ricardo G and others},
  journal={Astronomy \& Astrophysics},
  volume={664},
  pages={A18},
  year={2022},
  publisher={EDP Sciences}
}

@article{zhang2022bingo,
  title={The BINGO project-VI. H I halo occupation distribution and mock building},
  author={Zhang, Jiajun and Motta, Pablo and Novaes, Camila P and Abdalla, Filipe B and Costa, Andre A and Wang, Bin and Zhu, Zhenghao and Shan, Chenxi and Xu, Haiguang and Abdalla, Elcio and others},
  journal={Astronomy \& Astrophysics},
  volume={664},
  pages={A19},
  year={2022},
  publisher={EDP Sciences}
}

@article{costa2022bingo,
  title={The BINGO project-VII. Cosmological forecasts from 21 cm intensity mapping},
  author={Costa, Andre A and Landim, Ricardo G and Novaes, Camila P and Xiao, Linfeng and Ferreira, Elisa GM and Abdalla, Filipe B and Wang, Bin and Abdalla, Elcio and Battye, Richard A and Marins, Alessandro and others},
  journal={Astronomy \& Astrophysics},
  volume={664},
  pages={A20},
  year={2022},
  publisher={EDP Sciences}
}

@article{novaes2022bingo,
  title={The BINGO project-VIII. Recovering the BAO signal in HI intensity mapping simulations},
  author={Novaes, Camila P and Zhang, Jiajun and de Mericia, Eduardo J and Abdalla, Filipe B and Liccardo, Vincenzo and Wuensche, Carlos A and Delabrouille, Jacques and Remazeilles, Mathieu and Santos, Larissa and Landim, Ricardo G and others},
  journal={Astronomy \& Astrophysics},
  volume={666},
  pages={A83},
  year={2022},
  publisher={EDP Sciences}
}

@article{santos2015cosmology,
  title={Cosmology with a SKA HI intensity mapping survey},
  author={Santos, Mario G and Bull, Philip and Alonso, David and Camera, Stefano and Ferreira, Pedro G and Bernardi, Gianni and Maartens, Roy and Viel, Matteo and Villaescusa-Navarro, Francisco and Abdalla, Filipe B and others},
  journal={arXiv preprint arXiv:1501.03989},
  year={2015}
}

@article{nan2011five,
  title={The five-hundred-meter aperture spherical radio telescope (FAST) project},
  author={Nan, Rendong and Li, Di and Jin, Chengjin and Wang, Qiming and Zhu, Lichun and Zhu, Wenbai and Zhang, Haiyan and Yue, Youling and Qian, Lei},
  journal={International Journal of Modern Physics D},
  volume={20},
  number={06},
  pages={989--1024},
  year={2011},
  publisher={World Scientific}
}

@article{bigot2015hi,
  title={HI intensity mapping with FAST},
  author={Bigot-Sazy, Marie-Anne and Ma, Yin-Zhe and Battye, Richard A and Browne, Ian WA and Chen, Tianyue and Dickinson, Clive and Harper, Stuart and Maffei, Bruno and Olivari, Lucas C and Wilkinson, Peter N},
  journal={arXiv preprint arXiv:1511.03006},
  year={2015}
}

@article{santos2023bingo,
  title={The BINGO Project IX: Search for Fast Radio Bursts--A Forecast for the BINGO Interferometry System},
  author={Santos, Marcelo V dos and Landim, Ricardo G and Hoerning, Gabriel A and Abdalla, Filipe B and Queiroz, Amilcar and Abdalla, Elcio and Wuensche, Carlos A and Wang, Bin and Barosi, Luciano and Villela, Thyrso and others},
  journal={arXiv preprint arXiv:2308.06805},
  year={2023}
}

@inproceedings{chen2012tianlai,
  title={The Tianlai project: a 21cm cosmology experiment},
  author={Chen, Xuelei},
  booktitle={International Journal of Modern Physics: Conference Series},
  volume={12},
  pages={256--263},
  year={2012},
  organization={World Scientific}
}

@inproceedings{bandura2014canadian,
  title={Canadian hydrogen intensity mapping experiment (CHIME) pathfinder},
  author={Bandura, Kevin and Addison, Graeme E and Amiri, Mandana and Bond, J Richard and Campbell-Wilson, Duncan and Connor, Liam and Cliche, Jean-Fran{\c{c}}ois and Davis, Greg and Deng, Meiling and Denman, Nolan and others},
  booktitle={Ground-based and Airborne Telescopes V},
  volume={9145},
  pages={738--757},
  year={2014},
  organization={SPIE}
}

@article{rumelhart1986learning,
  title={Learning representations by back-propagating errors},
  author={Rumelhart, David E and Hinton, Geoffrey E and Williams, Ronald J},
  journal={nature},
  volume={323},
  number={6088},
  pages={533--536},
  year={1986},
  publisher={Nature Publishing Group UK London}
}

@article{hildebrandt2010photometric,
  title={Photometric redshifts in cosmology},
  author={Hildebrandt, H and Brusa, M and Ilbert, O and Capak, P and Salvato, M and others},
  journal={The Astrophysical Journal},
  volume={721},
  number={1},
  pages={109},
  year={2010},
  publisher={IOP Publishing}
}

@article{bolzonella2000photometric,
  title={Photometric redshifts: An artificial neural network approach},
  author={Bolzonella, M and Miralles, JM and Pell{\'o}, R},
  journal={Astronomy and Astrophysics},
  volume={363},
  pages={476--492},
  year={2000}
}

@article{carliles2010random,
  title={Random Forests for Photometric Redshifts},
  author={Carliles, Steven and Budav{\'a}ri, Tam{\'a}s and Heinis, St{\'e}phane and Szalay, Alexander S and van den Berg, MA and others},
  journal={The Astrophysical Journal},
  volume={721},
  number={1},
  pages={118},
  year={2010},
  publisher={IOP Publishing}
}

@article{cavuoti2015photometric,
  title={Photometric redshifts with the WIde-field MOon Photometric Survey (WMOPS) using ANNz2},
  author={Cavuoti, S and Brescia, M and Longo, G and Napolitano, NR},
  journal={Monthly Notices of the Royal Astronomical Society},
  volume={452},
  number={3},
  pages={3109--3122},
  year={2015},
  publisher={Oxford University Press}
}

@article{hoyle2016comparison,
  title={A comparison of photometric redshift methods for Dark Energy Survey data},
  author={Hoyle, B and Rau, M and Abdalla, FB and Amara, A and Hamaus, N and others},
  journal={Monthly Notices of the Royal Astronomical Society},
  volume={455},
  number={4},
  pages={3774--3791},
  year={2016},
  publisher={Oxford University Press}
}

@article{friedman2001greedy,
  title={Greedy function approximation: A gradient boosting machine},
  author={Friedman, Jerome H},
  journal={Annals of statistics},
  pages={1189--1232},
  year={2001},
  publisher={JSTOR}
}

@article{gerdes2010boosted,
  title={Boosted decision tree photometric redshifts in the Dark Energy Survey},
  author={Gerdes, DW and Dodelson, S and DePoy, D and Diehl, HT and Frieman, J and others},
  journal={Monthly Notices of the Royal Astronomical Society},
  volume={402},
  number={3},
  pages={1881--1890},
  year={2010},
  publisher={Oxford University Press}
}

@article{ball2007robust,
  title={A robust K-nearest neighbor method for photometric redshifts},
  author={Ball, Nicholas M and Brunner, Robert J and Myers, Adam D and Tsvetanov, Zlatan I and Djorgovski, SG},
  journal={The Astrophysical Journal},
  volume={663},
  number={2},
  pages={774},
  year={2007},
  publisher={IOP Publishing}
}

@article{richards2002spectroscopic,
  title={Spectroscopic Target Selection in the Sloan Digital Sky Survey: TheQuasar Sample},
  author={Richards, Gordon T and Fan, Xiaohui and Newberg, Heidi Jo and Strauss, Michael A and Vanden Berk, Daniel E and Schneider, Donald P and Yanny, Brian and Boucher, Adam and Burles, Scott and Frieman, Joshua A and others},
  journal={The Astronomical Journal},
  volume={123},
  number={6},
  pages={2945},
  year={2002},
  publisher={IOP Publishing}
}

@article{lahav2012annz,
  title={ANNz: Artificial Neural Networks for estimating photometric redshifts},
  author={Lahav, Ofer and Collister, Adrian A},
  journal={Astrophysics Source Code Library},
  pages={ascl--1209},
  year={2012}
}

@article{richards2006efficiency,
  author  = {Richards, Gordon T. and others},
  title   = {Efficient Photometric Selection of Quasars from the Sloan Digital Sky Survey. II. $\sim$1,000,000 Quasars from Data Release 6},
  journal = {Astron. J.},
  year    = {2006},
  volume  = {131},
  pages   = {2766-2787},
  doi     = {10.1086/503559}
}

@article{fan2001color,
  author  = {Fan, Xiaohui and others},
  title   = {A Survey of $z > 5.8$ Quasars in the Sloan Digital Sky Survey. I. Discovery of Three New Quasars and the Spatial Density of Luminous Quasars at $z \sim 6$},
  journal = {Astron. J.},
  year    = {2001},
  volume  = {121},
  pages   = {54-65},
  doi     = {10.1086/318033}
}

@article{fan1999evolution,
  author  = {Fan, Xiaohui and others},
  title   = {Evolution of the Ionizing Background and the Epoch of Reionization from the Spectra of $z \sim 6$ Quasars},
  journal = {Astrophys. J.},
  year    = {1999},
  volume  = {526},
  pages   = {57-67},
  doi     = {10.1086/307039}
}

@article{hoyle2015photometric,
  author  = {Hoyle, B. and Rau, M. M. and Paech, K. and Bonnett, C. and Seitz, S.},
  title   = {Machine learning photometric redshifts with random forests and Gaussian processes},
  journal = {Mon. Not. R. Astron. Soc.},
  year    = {2015},
  volume  = {452},
  number  = {4},
  pages   = {4183-4194},
  doi     = {10.1093/mnras/stv1557}
}

@article{baldry2002morphological,
  author = {Baldry, Ivan K. and Glazebrook, Karl and Brinkmann, J. and Ivezi{\'c}, {\v{Z}}eljko and Lupton, Robert H. and Nichol, Robert C. and Szalay, Alexander S.},
  title   = {Morphological Galaxy Classification in the Sloan Digital Sky Survey},
  journal = {Astrophys. J.},
  year    = {2002},
  volume  = {569},
  pages   = {582-594},
  doi     = {10.1086/339477}
}

@article{ahumada202016th,
  title={The 16th data release of the sloan digital sky surveys},
  author={Ahumada and others},
  journal={The Astrophysical Journal Supplement Series},
  volume={249},
  number={1},
  pages={3},
  year={2020},
  publisher={IOP Publishing}
}

@article{vandenberk2001composite,
  author  = {Vanden Berk, Daniel E. and others},
  title   = {Composite Quasar Spectra from the Sloan Digital Sky Survey},
  journal = {Astron. J.},
  year    = {2001},
  volume  = {122},
  pages   = {549-564},
  doi     = {10.1086/321167}
}

@article{Richards2002,
  author = {Richards, Gordon T. and Fan, Xiaohui and Schneider, Donald P. and Vanden Berk, Daniel E. and Strauss, Michael A. and York, Donald G. and Anderson, Scott F. and Anderson, Jr., John E. and Annis, James and Bahcall, Neta A. and others},
  title = {Colors of 2625 Quasars at 0 < z < 5 Measured in the Sloan Digital Sky Survey Photometric System},
  journal = {The Astronomical Journal},
  volume = {123},
  number = {6},
  pages = {2945},
  year = {2002},
  doi = {10.1086/340187},
}

@article{abdalla2025machine,
  title={Machine learning analysis of Photometric data from the Dark Energy Survey},
  author={Abdalla, Elcio and Abdalla, Filipe B and Marins, Alessandro and Queiroz, Amilcar and Ribeiro, Rafael M and Souza, Alex SC},
  journal={arXiv preprint arXiv:2508.10191},
  year={2025}
}

@article{Weiner2005,
  author = {Weiner, Benjamin J. and Phillips, A. C. and Faber, S. M. and Willmer, C. N. A. and Vogt, N. P. and Simard, L. and Gebhardt, K. and Im, M. and Koo, D. C. and Sarajedini, V. L. and others},
  title = {A Spectroscopic Survey of Redshift 1.4 Galaxies in the GOODS-North Field: The Redshift Catalog},
  journal = {The Astrophysical Journal},
  volume = {620},
  number = {2},
  pages = {595},
  year = {2005},
  doi = {10.1086/431416},
}

@article{Skrzypek2016,
  author = {Skrzypek, N. and Warren, S. J. and Faherty, J. K.},
  title = {UKIDSS counterparts to cool WISE-selected quasars: revealing a population of M-dwarf/quasar misidentifications},
  journal = {Monthly Notices of the Royal Astronomical Society},
  volume = {458},
  number = {3},
  pages = {2971–2977},
  year = {2016},
  doi = {10.1093/mnras/stv2932},
}

@article{dawson2013sdss,
  author  = {Dawson, Kyle S. and others},
  title   = {The Baryon Oscillation Spectroscopic Survey of SDSS-III},
  journal = {Astron. J.},
  year    = {2013},
  volume  = {145},
  pages   = {10},
  doi     = {10.1088/0004-6256/145/1/10}
}

@article{abbott2018dark,
  title={The Dark Energy Survey: Data Release 1},
  author={Abbott, T.M.C. and others},
  journal={The Astrophysical Journal Supplement Series},
  volume={239},
  number={2},
  pages={18},
  year={2018},
  doi={10.3847/1538-4365/aae9f0}
}

@article{abbott2021dark,
  title={Dark Energy Survey Year 3 Results: Data Release 2},
  author={Abbott, T.M.C. and others},
  journal={The Astrophysical Journal Supplement Series},
  volume={255},
  number={2},
  pages={20},
  year={2021},
  doi={10.3847/1538-4365/ac00b3}
}

@article{bolton2012spectral,
  title={Spectral classification and redshift measurement for the SDSS-III baryon oscillation spectroscopic survey},
  author={Bolton, Adam S and Schlegel, David J and Aubourg, {\'E}ric and Bailey, Stephen and Bhardwaj, Vaishali and Brownstein, Joel R and Burles, Scott and Chen, Yan-Mei and Dawson, Kyle and Eisenstein, Daniel J and others},
  journal={The Astronomical Journal},
  volume={144},
  number={5},
  pages={144},
  year={2012},
  publisher={IOP Publishing}
}

@inproceedings{kunszt2001hierarchical,
  title={The hierarchical triangular mesh},
  author={Kunszt, Peter Z and Szalay, Alexander S and Thakar, Aniruddha R},
  booktitle={Mining the Sky},
  pages={631--637},
  year={2001},
  publisher={Springer}
}

@article{to2021dark,
  title = {Dark Energy Survey Year 3 results: Cosmological constraints from galaxy clustering and weak lensing},
  author = {Abbott, T. M. C. and Aguena, M. and Alarcon, A. and Allam, S. and Alves, O. and Amon, A. and Andrade-Oliveira, F. and Annis, J. and Avila, S. and Bacon, D. and Baxter, E. and Bechtol, K. and Becker, M. R. and Bernstein, G. M. and Bhargava, S. and Birrer, S. and Blazek, J. and Brandao-Souza, A. and Bridle, S. L. and Brooks, D. and Buckley-Geer, E. and Burke, D. L. and Camacho, H. and Campos, A. and Carnero Rosell, A. and Carrasco Kind, M. and Carretero, J. and Castander, F. J. and Cawthon, R. and Chang, C. and Chen, A. and Chen, R. and Choi, A. and Conselice, C. and Cordero, J. and Costanzi, M. and Crocce, M. and da Costa, L. N. and da Silva Pereira, M. E. and Davis, C. and Davis, T. M. and De Vicente, J. and DeRose, J. and Desai, S. and Di Valentino, E. and Diehl, H. T. and Dietrich, J. P. and Dodelson, S. and Doel, P. and Doux, C. and Drlica-Wagner, A. and Eckert, K. and Eifler, T. F. and Elsner, F. and Elvin-Poole, J. and Everett, S. and Evrard, A. E. and Fang, X. and Farahi, A. and Fernandez, E. and Ferrero, I. and Fert\'e, A. and Fosalba, P. and Friedrich, O. and Frieman, J. and Garc\'{\i}a-Bellido, J. and Gatti, M. and Gaztanaga, E. and Gerdes, D. W. and Giannantonio, T. and Giannini, G. and Gruen, D. and Gruendl, R. A. and Gschwend, J. and Gutierrez, G. and Harrison, I. and Hartley, W. G. and Herner, K. and Hinton, S. R. and Hollowood, D. L. and Honscheid, K. and Hoyle, B. and Huff, E. M. and Huterer, D. and Jain, B. and James, D. J. and Jarvis, M. and Jeffrey, N. and Jeltema, T. and Kovacs, A. and Krause, E. and Kron, R. and Kuehn, K. and Kuropatkin, N. and Lahav, O. and Leget, P.-F. and Lemos, P. and Liddle, A. R. and Lidman, C. and Lima, M. and Lin, H. and MacCrann, N. and Maia, M. A. G. and Marshall, J. L. and Martini, P. and McCullough, J. and Melchior, P. and Mena-Fern\'andez, J. and Menanteau, F. and Miquel, R. and Mohr, J. J. and Morgan, R. and Muir, J. and Myles, J. and Nadathur, S. and Navarro-Alsina, A. and Nichol, R. C. and Ogando, R. L. C. and Omori, Y. and Palmese, A. and Pandey, S. and Park, Y. and Paz-Chinch\'on, F. and Petravick, D. and Pieres, A. and Plazas Malag\'on, A. A. and Porredon, A. and Prat, J. and Raveri, M. and Rodriguez-Monroy, M. and Rollins, R. P. and Romer, A. K. and Roodman, A. and Rosenfeld, R. and Ross, A. J. and Rykoff, E. S. and Samuroff, S. and S\'anchez, C. and Sanchez, E. and Sanchez, J. and Sanchez Cid, D. and Scarpine, V. and Schubnell, M. and Scolnic, D. and Secco, L. F. and Serrano, S. and Sevilla-Noarbe, I. and Sheldon, E. and Shin, T. and Smith, M. and Soares-Santos, M. and Suchyta, E. and Swanson, M. E. C. and Tabbutt, M. and Tarle, G. and Thomas, D. and To, C. and Troja, A. and Troxel, M. A. and Tucker, D. L. and Tutusaus, I. and Varga, T. N. and Walker, A. R. and Weaverdyck, N. and Wechsler, R. and Weller, J. and Yanny, B. and Yin, B. and Zhang, Y. and Zuntz, J.},
  collaboration = {DES Collaboration},
  journal = {Phys. Rev. D},
  volume = {105},
  issue = {2},
  pages = {023520},
  numpages = {42},
  year = {2022},
  month = {Jan},
  publisher = {American Physical Society},
  doi = {10.1103/PhysRevD.105.023520},
  url = {https://link.aps.org/doi/10.1103/PhysRevD.105.023520}
}

@inproceedings{descollaboration2005,
  title={The Dark Energy Survey},
  author={The Dark Energy Survey Collaboration},
  booktitle={The Dark Energy Survey White Paper},
  year={2005}
}

@article{kessler2015results,
  title={Results from the Dark Energy Survey Supernova Program},
  author={Kessler, R. and others},
  journal={The Astronomical Journal},
  volume={150},
  number={6},
  pages={172},
  year={2015},
  doi={10.1088/0004-6256/150/6/172}
}

@article{des2016,  
  author={Dark Energy Survey Collaboration},  
  title={The Dark Energy Survey: More than dark energy – an overview},  
  journal={MNRAS},  
  year={2016},  
  volume={460},  
  pages={1270}  
}

@article{anzu2018,  
  author={Sadeh, I. and others},  
  title={ANNz2: Photometric Redshift and Probability Distribution Function Estimation using Machine Learning},  
  journal={ApJS},  
  year={2016},  
  volume={219},  
  pages={1}  
}

@article{fan2023quasars,
  title={Quasars and the intergalactic medium at cosmic dawn},
  author={Fan, Xiaohui and Ba{\~n}ados, Eduardo and Simcoe, Robert A},
  journal={Annual Review of Astronomy and Astrophysics},
  volume={61},
  number={1},
  pages={373--426},
  year={2023},
  publisher={Annual Reviews}
}

@article{van2010shear,
  title={Shear and magnification: cosmic complementarity},
  author={Van Waerbeke, L},
  journal={Monthly Notices of the Royal Astronomical Society},
  volume={401},
  number={3},
  pages={2093--2100},
  year={2010},
  publisher={Blackwell Publishing Ltd Oxford, UK}
}

@article{cover1967nearest,
  title={Nearest neighbor pattern classification},
  author={Cover, Thomas and Hart, Peter},
  journal={IEEE Transactions on Information Theory},
  volume={13},
  number={1},
  pages={21--27},
  year={1967}
}

@article{viquar2019machine,
  title={Machine learning in astronomy: A case study in quasar-star classification},
  author={Viquar, Mohammed and Basak, Suryoday and Dasgupta, Ariruna and Agrawal, Surbhi and Saha, Snehanshu},
  journal={Emerging Technologies in Data Mining and Information Security: Proceedings of IEMIS 2018, Volume 3},
  pages={827--836},
  year={2019},
  publisher={Springer}
}

@article{li2008k,
  title={k-Nearest Neighbors for automated classification of celestial objects},
  author={Li, LiLi and Zhang, YanXia and Zhao, YongHeng},
  journal={Science in China Series G: Physics, Mechanics and Astronomy},
  volume={51},
  number={7},
  pages={916--922},
  year={2008},
  publisher={Springer}
}

@article{van2008visualizing,
  title={Visualizing data using t-SNE.},
  author={Van der Maaten, Laurens and Hinton, Geoffrey},
  journal={Journal of machine learning research},
  volume={9},
  number={11},
  year={2008}
}

@article{zheng2020multiple,
  title={Multiple measurements of quasars acting as standard probes: exploring the cosmic distance duality relation at higher redshift},
  author={Zheng, Xiaogang and Liao, Kai and Biesiada, Marek and Cao, Shuo and Liu, Tong-Hua and Zhu, Zong-Hong},
  journal={The Astrophysical Journal},
  volume={892},
  number={2},
  pages={103},
  year={2020},
  publisher={IOP Publishing}
}

@article{zhang2023trinity,
  title={Trinity II: The luminosity-dependent bias of the supermassive black hole mass--galaxy mass relation for bright quasars at z= 6},
  author={Zhang, Haowen and Behroozi, Peter and Volonteri, Marta and Silk, Joseph and Fan, Xiaohui and Aird, James and Yang, Jinyi and Hopkins, Philip F},
  journal={Monthly Notices of the Royal Astronomical Society: Letters},
  volume={523},
  number={1},
  pages={L69--L74},
  year={2023},
  publisher={Oxford University Press}
}

@article{wolpert1992stacked,
  title={Stacked generalization},
  author={Wolpert, David H.},
  journal={Neural networks},
  volume={5},
  number={2},
  pages={241--259},
  year={1992},
  publisher={Elsevier}
}

@article{ivezic2019lsst,
  author  = {Ivezi{\'c}, {\v{Z}}eljko and others},
  title   = {LSST: From Science Drivers to Reference Design and Anticipated Data Products},
  journal = {Astrophys. J.},
  year    = {2019},
  volume  = {873},
  pages   = {111},
  doi     = {10.3847/1538-4357/ab042c}
}

@article{laureijs2011euclid,
  author  = {Laureijs, René and others},
  title   = {Euclid Definition Study Report},
  journal = {arXiv:1110.3193},
  year    = {2011}
}

@article{rees1984black,
  author  = {Rees, Martin J.},
  title   = {Black Hole Models for Active Galactic Nuclei},
  journal = {Annu. Rev. Astron. Astrophys.},
  year    = {1984},
  volume  = {22},
  pages   = {471-506},
  doi     = {10.1146/annurev.aa.22.090184.002351}
}

@article{richards2002color,
  author  = {Richards, Gordon T. and others},
  title   = {Colors of 2625 Quasars at 0 < z < 5 Measured in the Sloan Digital Sky Survey Photometric System},
  journal = {Astron. J.},
  year    = {2002},
  volume  = {123},
  pages   = {2945},
  doi     = {10.1086/340187}
}

@article{eisenstein2005detection,
  author  = {Eisenstein, Daniel J. and others},
  title   = {Detection of the Baryon Acoustic Peak in the Large-Scale Correlation Function of SDSS Luminous Red Galaxies},
  journal = {Astrophys. J.},
  year    = {2005},
  volume  = {633},
  pages   = {560},
  doi     = {10.1086/466512}
}

@article{rauch1998lyman,
  author  = {Rauch, Michael and others},
  title   = {The Lyman Alpha Forest in the Spectra of Quasars},
  journal = {Astrophys. J.},
  year    = {1998},
  volume  = {489},
  pages   = {7},
  doi     = {10.1086/304753}
}

@article{york2000sdss,
  author  = {York, Donald G. and others},
  title   = {The Sloan Digital Sky Survey: Technical Summary},
  journal = {Astron. J.},
  year    = {2000},
  volume  = {120},
  pages   = {1579},
  doi     = {10.1086/301513}
}

@article{pedregosa2011scikit,
  author  = {Pedregosa, Fabian and others},
  title   = {Scikit-learn: Machine Learning in Python},
  journal = {J. Mach. Learn. Res.},
  year    = {2011},
  volume  = {12},
  pages   = {2825–2830}
}

@article{tanaka2018photometric,
  title={Photometric redshifts for Hyper Suprime-Cam Subaru strategic program data release 1},
  author={Tanaka, Masayuki and Coupon, Jean and Hsieh, Bau-Ching and Mineo, Sogo and Nishizawa, Atsushi J and Speagle, Joshua and Furusawa, Hisanori and Miyazaki, Satoshi and Murayama, Hitoshi},
  journal={Publications of the Astronomical Society of Japan},
  volume={70},
  number={SP1},
  pages={S9},
  year={2018},
  publisher={Oxford University Press}
}

@article{kaiser2010panstarrs,
  author  = {Kaiser, Nicholas and others},
  title   = {The Pan-STARRS Wide-field Optical/NIR Imaging Survey},
  journal = {Proc. SPIE},
  year    = {2010},
  volume  = {7733},
  doi     = {10.1117/12.859188}
}

@inproceedings{bottou2010large,
  title={Large-scale machine learning with stochastic gradient descent},
  author={Bottou, L{\'e}on},
  booktitle={Proceedings of COMPSTAT'2010},
  pages={177--186},
  year={2010},
  publisher={Springer},
  doi={10.1007/978-3-7908-2604-3_16}
}

@article{beck2017photometric,
  author  = {Beck, Róbert and others},
  title   = {Photometric Redshift Estimation with a Convolutional Neural Network},
  journal = {Mon. Not. R. Astron. Soc.},
  year    = {2017},
  volume  = {472},
  pages   = {949},
  doi     = {10.1093/mnras/stx1907}
}

@article{peters2015quasar,
  title={Quasar classification using color and variability},
  author={Peters, Christina M and Richards, Gordon T and Myers, Adam D and Strauss, Michael A and Schmidt, Kasper B and Ivezic, {\v{Z}}eljko and Ross, Nicholas P and MacLeod, Chelsea L and Riegel, Ryan},
  journal={The Astrophysical Journal},
  volume={811},
  number={2},
  pages={95},
  year={2015},
  publisher={IOP Publishing}
}

@article{repp2016systematic,
  title={A systematic search for lensed high-redshift galaxies in HST images of MACS clusters},
  author={Repp, Andrew and Ebeling, Harald and Richard, Johan},
  journal={Monthly Notices of the Royal Astronomical Society},
  volume={457},
  number={2},
  pages={1399--1409},
  year={2016},
  publisher={Oxford University Press}
}

@article{wu2004color,
  title={Color-redshift relations and photometric redshift estimations of quasars in large sky surveys},
  author={Wu, Xue-Bing and Zhang, Wei and Zhou, Xu},
  journal={Chinese Journal of Astronomy and Astrophysics},
  volume={4},
  number={1},
  pages={17},
  year={2004},
  publisher={IOP Publishing}
}

@article{han2016photometric,
  title={Photometric redshift estimation for quasars by integration of KNN and SVM},
  author={Han, Bo and Ding, Hong-Peng and Zhang, Yan-Xia and Zhao, Yong-Heng},
  journal={Research in Astronomy and Astrophysics},
  volume={16},
  number={5},
  pages={005},
  year={2016},
  publisher={IOP Publishing}
}

@article{yang2017quasar,
  title={Quasar photometric redshifts and candidate selection: A new algorithm based on optical and mid-infrared photometric data},
  author={Yang, Qian and Wu, Xue-Bing and Fan, Xiaohui and Jiang, Linhua and McGreer, Ian and Green, Richard and Yang, Jinyi and Schindler, Jan-Torge and Wang, Feige and Zuo, Wenwen and others},
  journal={The Astronomical Journal},
  volume={154},
  number={6},
  pages={269},
  year={2017},
  publisher={IOP Publishing}
}

@article{Frieman_2008,
   title={Dark Energy and the Accelerating Universe},
   volume={46},
   ISSN={1545-4282},
   url={http://dx.doi.org/10.1146/annurev.astro.46.060407.145243},
   DOI={10.1146/annurev.astro.46.060407.145243},
   number={1},
   journal={Annual Review of Astronomy and Astrophysics},
   publisher={Annual Reviews},
   author={Frieman, Joshua A. and Turner, Michael S. and Huterer, Dragan},
   year={2008},
   month=sep, pages={385–432} }

@article{Amendola_2013,
   title={Cosmology and Fundamental Physics with the Euclid Satellite},
   volume={16},
   ISSN={1433-8351},
   url={http://dx.doi.org/10.12942/lrr-2013-6},
   DOI={10.12942/lrr-2013-6},
   number={1},
   journal={Living Reviews in Relativity},
   publisher={Springer Science and Business Media LLC},
   author={Amendola, Luca and Appleby, Stephen and Bacon, David and Baker, Tessa and Baldi, Marco and Bartolo, Nicola and Blanchard, Alain and Bonvin, Camille and Borgani, Stefano and Branchini, Enzo and Burrage, Clare and Camera, Stefano and Carbone, Carmelita and Casarini, Luciano and Cropper, Mark and de Rham, Claudia and Di Porto, Cinzia and Ealet, Anne and Ferreira, Pedro G. and Finelli, Fabio and García-Bellido, Juan and Giannantonio, Tommaso and Guzzo, Luigi and Heavens, Alan and Heisenberg, Lavinia and Heymans, Catherine and Hoekstra, Henk and Hollenstein, Lukas and Holmes, Rory and Horst, Ole and Jahnke, Knud and Kitching, Thomas D. and Koivisto, Tomi and Kunz, Martin and La Vacca, Giuseppe and March, Marisa and Majerotto, Elisabetta and Markovic, Katarina and Marsh, David and Marulli, Federico and Massey, Richard and Mellier, Yannick and Mota, David F. and Nunes, Nelson J. and Percival, Will and Pettorino, Valeria and Porciani, Cristiano and Quercellini, Claudia and Read, Justin and Rinaldi, Massimiliano and Sapone, Domenico and Scaramella, Roberto and Skordis, Constantinos and Simpson, Fergus and Taylor, Andy and Thomas, Shaun and Trotta, Roberto and Verde, Licia and Vernizzi, Filippo and Vollmer, Adrian and Wang, Yun and Weller, Jochen and Zlosnik, Tom},
   year={2013},
   month=sep }

@article{Cole_2005,
   title={The 2dF Galaxy Redshift Survey: power-spectrum analysis of the final data set and cosmological implications},
   volume={362},
   ISSN={1365-2966},
   url={http://dx.doi.org/10.1111/j.1365-2966.2005.09318.x},
   DOI={10.1111/j.1365-2966.2005.09318.x},
   number={2},
   journal={Monthly Notices of the Royal Astronomical Society},
   publisher={Oxford University Press (OUP)},
   author={Cole, Shaun and Percival, Will J. and Peacock, John A. and Norberg, Peder and Baugh, Carlton M. and Frenk, Carlos S. and Baldry, Ivan and Bland-Hawthorn, Joss and Bridges, Terry and Cannon, Russell and Colless, Matthew and Collins, Chris and Couch, Warrick and Cross, Nicholas J. G. and Dalton, Gavin and Eke, Vincent R. and De Propris, Roberto and Driver, Simon P. and Efstathiou, George and Ellis, Richard S. and Glazebrook, Karl and Jackson, Carole and Jenkins, Adrian and Lahav, Ofer and Lewis, Ian and Lumsden, Stuart and Maddox, Steve and Madgwick, Darren and Peterson, Bruce A. and Sutherland, Will and Taylor, Keith},
   year={2005},
   month=sep, pages={505–534} }

@article{Alam_2017,
   title={The clustering of galaxies in the completed SDSS-III Baryon Oscillation Spectroscopic Survey: cosmological analysis of the DR12 galaxy sample},
   volume={470},
   ISSN={1365-2966},
   url={http://dx.doi.org/10.1093/mnras/stx721},
   DOI={10.1093/mnras/stx721},
   number={3},
   journal={Monthly Notices of the Royal Astronomical Society},
   publisher={Oxford University Press (OUP)},
   author={Alam, Shadab and Ata, Metin and Bailey, Stephen and Beutler, Florian and Bizyaev, Dmitry and Blazek, Jonathan A. and Bolton, Adam S. and Brownstein, Joel R. and Burden, Angela and Chuang, Chia-Hsun and Comparat, Johan and Cuesta, Antonio J. and Dawson, Kyle S. and Eisenstein, Daniel J. and Escoffier, Stephanie and Gil-Marín, Héctor and Grieb, Jan Niklas and Hand, Nick and Ho, Shirley and Kinemuchi, Karen and Kirkby, David and Kitaura, Francisco and Malanushenko, Elena and Malanushenko, Viktor and Maraston, Claudia and McBride, Cameron K. and Nichol, Robert C. and Olmstead, Matthew D. and Oravetz, Daniel and Padmanabhan, Nikhil and Palanque-Delabrouille, Nathalie and Pan, Kaike and Pellejero-Ibanez, Marcos and Percival, Will J. and Petitjean, Patrick and Prada, Francisco and Price-Whelan, Adrian M. and Reid, Beth A. and Rodríguez-Torres, Sergio A. and Roe, Natalie A. and Ross, Ashley J. and Ross, Nicholas P. and Rossi, Graziano and Rubiño-Martín, Jose Alberto and Saito, Shun and Salazar-Albornoz, Salvador and Samushia, Lado and Sánchez, Ariel G. and Satpathy, Siddharth and Schlegel, David J. and Schneider, Donald P. and Scóccola, Claudia G. and Seo, Hee-Jong and Sheldon, Erin S. and Simmons, Audrey and Slosar, Anže and Strauss, Michael A. and Swanson, Molly E. C. and Thomas, Daniel and Tinker, Jeremy L. and Tojeiro, Rita and Magaña, Mariana Vargas and Vazquez, Jose Alberto and Verde, Licia and Wake, David A. and Wang, Yuting and Weinberg, David H. and White, Martin and Wood-Vasey, W. Michael and Yèche, Christophe and Zehavi, Idit and Zhai, Zhongxu and Zhao, Gong-Bo},
   year={2017},
   month=mar, pages={2617–2652} }

@article{Anderson_2014,
   title={The clustering of galaxies in the SDSS-III Baryon Oscillation Spectroscopic Survey: baryon acoustic oscillations in the Data Releases 10 and 11 Galaxy samples},
   volume={441},
   ISSN={0035-8711},
   url={http://dx.doi.org/10.1093/mnras/stu523},
   DOI={10.1093/mnras/stu523},
   number={1},
   journal={Monthly Notices of the Royal Astronomical Society},
   publisher={Oxford University Press (OUP)},
   author={Anderson, Lauren and Aubourg, {\'E}ric and Bailey, Stephen and Beutler, Florian and Bhardwaj, Vaishali and Blanton, Michael and Bolton, Adam S. and Brinkmann, J. and Brownstein, Joel R. and Burden, Angela and Chuang, Chia-Hsun and Cuesta, Antonio J. and Dawson, Kyle S. and Eisenstein, Daniel J. and Escoffier, Stephanie and Gunn, James E. and Guo, Hong and Ho, Shirley and Honscheid, Klaus and Howlett, Cullan and Kirkby, David and Lupton, Robert H. and Manera, Marc and Maraston, Claudia and McBride, Cameron K. and Mena, Olga and Montesano, Francesco and Nichol, Robert C. and Nuza, Sebasti{\'a}n E. and Olmstead, Matthew D. and Padmanabhan, Nikhil and Palanque-Delabrouille, Nathalie and Parejko, John and Percival, Will J. and Petitjean, Patrick and Prada, Francisco and Price-Whelan, Adrian M. and Reid, Beth and Roe, Natalie A. and Ross, Ashley J. and Ross, Nicholas P. and Sabiu, Cristiano G. and Saito, Shun and Samushia, Lado and S{\'a}nchez, Ariel G. and Schlegel, David J. and Schneider, Donald P. and Scoccola, Claudia G. and Seo, Hee-Jong and Skibba, Ramin A. and Strauss, Michael A. and Swanson, Molly E. C. and Thomas, Daniel and Tinker, Jeremy L. and Tojeiro, Rita and Maga{\~n}a, Mariana Vargas and Verde, Licia and Wake, David A. and Weaver, Benjamin A. and Weinberg, David H. and White, Martin and Xu, Xiaoying and Y{\`e}che, Christophe and Zehavi, Idit and Zhao, Gong-Bo},
   year={2014},
   month=apr, pages={24–62} }

@article{Oyaizu_2008,
   title={A Galaxy Photometric Redshift Catalog for the Sloan Digital Sky Survey Data Release 6},
   volume={674},
   ISSN={1538-4357},
   url={http://dx.doi.org/10.1086/523666},
   DOI={10.1086/523666},
   number={2},
   journal={The Astrophysical Journal},
   publisher={American Astronomical Society},
   author={Oyaizu, Hiroaki and Lima, Marcos and Cunha, Carlos E. and Lin, Huan and Frieman, Joshua and Sheldon, Erin S.},
   year={2008},
   month=feb, pages={768–783} }

@article{almosallam2016gpz,
  title={GPz: non-stationary sparse Gaussian processes for heteroscedastic uncertainty estimation in photometric redshifts},
  author={Almosallam, Ibrahim A and Jarvis, Matt J and Roberts, Stephen J},
  journal={Monthly Notices of the Royal Astronomical Society},
  volume={462},
  number={1},
  pages={726--739},
  year={2016},
  publisher={The Royal Astronomical Society}
}

@article{almosallam2016sparse,
  title={A sparse Gaussian process framework for photometric redshift estimation},
  author={Almosallam, Ibrahim A and Lindsay, Sam N and Jarvis, Matt J and Roberts, Stephen J},
  journal={Monthly Notices of the Royal Astronomical Society},
  volume={455},
  number={3},
  pages={2387--2401},
  year={2016},
  publisher={Oxford University Press}
}

@article{rivera2018degradation,
  title={Degradation analysis in the estimation of photometric redshifts from non-representative training sets},
  author={Rivera, JD and Moraes, B and Merson, AI and Jouvel, S and Abdalla, FB and Abdalla, MCB},
  journal={Monthly Notices of the Royal Astronomical Society},
  volume={477},
  number={4},
  pages={4330--4347},
  year={2018},
  publisher={Oxford University Press}
}

@article{arnouts2011lephare,
  title={Lephare: Photometric analysis for redshift estimate},
  author={Arnouts, S and Ilbert, O},
  journal={Astrophysics Source Code Library},
  pages={ascl--1108},
  year={2011}
}

@article{joudaki2018kids,
  title={KiDS-450+ 2dFLenS: Cosmological parameter constraints from weak gravitational lensing tomography and overlapping redshift-space galaxy clustering},
  author={Joudaki, Shahab and Blake, Chris and Johnson, Andrew and Amon, Alexandra and Asgari, Marika and Choi, Ami and Erben, Thomas and Glazebrook, Karl and Harnois-D{\'e}raps, Joachim and Heymans, Catherine and others},
  journal={Monthly Notices of the Royal Astronomical Society},
  volume={474},
  number={4},
  pages={4894--4924},
  year={2018},
  publisher={Oxford University Press}
}

@article{prat2018dark,
  title={Dark Energy Survey year 1 results: Galaxy-galaxy lensing},
  author={Prat, Judit and S{\'a}nchez, C and Fang, Y and Gruen, D and Elvin-Poole, J and Kokron, N and Secco, LF and Jain, B and Miquel, R and MacCrann, N and others},
  journal={Physical Review D},
  volume={98},
  number={4},
  pages={042005},
  year={2018},
  publisher={APS}
}

@article{hildebrandt2017kids,
  title={KiDS-450: cosmological parameter constraints from tomographic weak gravitational lensing},
  author={Hildebrandt, H and Viola, M and Heymans, C and Joudaki, Shahab and Kuijken, K and Blake, Chris and Erben, T and Joachimi, B and Klaes, D and Miller, L t and others},
  journal={Monthly Notices of the Royal Astronomical Society},
  volume={465},
  number={2},
  pages={1454--1498},
  year={2017},
  publisher={Oxford University Press}
}

@article{benitez2000bayesian,
  title={Bayesian photometric redshift estimation},
  author={Benitez, Narciso},
  journal={The Astrophysical Journal},
  volume={536},
  number={2},
  pages={571--583},
  year={2000}
}

@article{PhysRevD.100.043501,
  title = {Dark Energy Survey Year 1 Results: Tomographic cross-correlations between Dark Energy Survey galaxies and CMB lensing from South Pole $\mathrm{Telescope}+\mathit{Planck}$},
  author = {Omori, Y. and Giannantonio, T. and Porredon, A. and Baxter, E. J. and Chang, C. and Crocce, M. and Fosalba, P. and Alarcon, A. and Banik, N. and Blazek, J. and Bleem, L. E. and Bridle, S. L. and Cawthon, R. and Choi, A. and Chown, R. and Crawford, T. and Dodelson, S. and Drlica-Wagner, A. and Eifler, T. F. and Elvin-Poole, J. and Friedrich, O. and Gruen, D. and Holder, G. P. and Huterer, D. and Jain, B. and Jarvis, M. and Kirk, D. and Kokron, N. and Krause, E. and MacCrann, N. and Muir, J. and Prat, J. and Reichardt, C. L. and Ross, A. J. and Rozo, E. and Rykoff, E. S. and S\'anchez, C. and Secco, L. F. and Simard, G. and Wechsler, R. H. and Zuntz, J. and Abbott, T. M. C. and Abdalla, F. B. and Allam, S. and Avila, S. and Aylor, K. and Benson, B. A. and Bernstein, G. M. and Bertin, E. and Bianchini, F. and Brooks, D. and Buckley-Geer, E. and Burke, D. L. and Carlstrom, J. E. and Carnero Rosell, A. and Carrasco Kind, M. and Carretero, J. and Castander, F. J. and Chang, C. L. and Cho, H-M. and Crites, A. T. and Cunha, C. E. and da Costa, L. N. and de Haan, T. and Davis, C. and De Vicente, J. and Desai, S. and Diehl, H. T. and Dietrich, J. P. and Dobbs, M. A. and Everett, W. B. and Doel, P. and Estrada, J. and Flaugher, B. and Frieman, J. and Garc\'{\i}a-Bellido, J. and Gaztanaga, E. and Gerdes, D. W. and George, E. M. and Gruendl, R. A. and Gschwend, J. and Gutierrez, G. and Halverson, N. W. and Harrington, N. L. and Hartley, W. G. and Hollowood, D. L. and Holzapfel, W. L. and Honscheid, K. and Hou, Z. and Hoyle, B. and Hrubes, J. D. and James, D. J. and Jeltema, T. and Kuehn, K. and Kuropatkin, N. and Lee, A. T. and Leitch, E. M. and Lima, M. and Luong-Van, D. and Manzotti, A. and Marrone, D. P. and Marshall, J. L. and McMahon, J. J. and Melchior, P. and Menanteau, F. and Meyer, S. S. and Miller, C. J. and Miquel, R. and Mocanu, L. M. and Mohr, J. J. and Natoli, T. and Padin, S. and Plazas, A. A. and Pryke, C. and Romer, A. K. and Roodman, A. and Ruhl, J. E. and Sanchez, E. and Scarpine, V. and Schaffer, K. K. and Schubnell, M. and Serrano, S. and Sevilla-Noarbe, I. and Shirokoff, E. and Smith, M. and Soares-Santos, M. and Sobreira, F. and Staniszewski, Z. and Stark, A. A. and Story, K. T. and Suchyta, E. and Swanson, M. E. C. and Tarle, G. and Thomas, D. and Troxel, M. A. and Vanderlinde, K. and Vieira, J. D. and Walker, A. R. and Wu, W. L. K. and Zahn, O.},
  collaboration = {DES and SPT Collaborations},
  journal = {Phys. Rev. D},
  volume = {100},
  issue = {4},
  pages = {043501},
  numpages = {18},
  year = {2019},
  month = {Aug},
  publisher = {American Physical Society},
  doi = {10.1103/PhysRevD.100.043501},
  url = {https://link.aps.org/doi/10.1103/PhysRevD.100.043501}
}

@article{huetsi2006acoustic,
  title={Acoustic oscillations in the SDSS DR4 luminous red galaxy sample power spectrum},
  author={Huetsi, Gert},
  journal={Astronomy \& Astrophysics},
  volume={449},
  number={3},
  pages={891--902},
  year={2006},
  publisher={EDP Sciences}
}

@article{ilbert2006accurate,
  title={Accurate photometric redshifts for the CFHT legacy survey calibrated using the VIMOS VLT deep survey},
  author={Ilbert, Olivier and Arnouts, S and Mccracken, Henry J and Bolzonella, M and Bertin, Emmanuel and Le F{\`e}vre, Olivier and Mellier, Yannick and Zamorani, G and Pello, R and Iovino, Angela and others},
  journal={Astronomy \& Astrophysics},
  volume={457},
  number={3},
  pages={841--856},
  year={2006},
  publisher={EDP Sciences}
}

@article{newman2013deep2,
  title={The DEEP2 galaxy redshift survey: Design, observations, data reduction, and redshifts},
  author={Newman, Jeffrey A and Cooper, Michael C and Davis, Marc and Faber, SM and Coil, Alison L and Guhathakurta, Puragra and Koo, David C and Phillips, Andrew C and Conroy, Charlie and Dutton, Aaron A and others},
  journal={The Astrophysical Journal Supplement Series},
  volume={208},
  number={1},
  pages={5},
  year={2013},
  publisher={IOP Publishing}
}

@article{masters2019complete,
  title={The complete calibration of the color--redshift relation (C3R2) survey: analysis and data release 2},
  author={Masters, Daniel C and Stern, Daniel K and Cohen, Judith G and Capak, Peter L and Stanford, S Adam and Hernitschek, Nina and Galametz, Audrey and Davidzon, Iary and Rhodes, Jason D and Sanders, Dave and others},
  journal={The Astrophysical Journal},
  volume={877},
  number={2},
  pages={81},
  year={2019},
  publisher={IOP Publishing}
}

@article{bovy2012photometric,
  title={Photometric redshifts and quasar probabilities from a single, data-driven generative model},
  author={Bovy, Jo and Myers, Adam D and Hennawi, Joseph F and Hogg, David W and McMahon, Richard G and Schiminovich, David and Sheldon, Erin S and Brinkmann, Jon and Schneider, Donald P and Weaver, Benjamin A},
  journal={The astrophysical journal},
  volume={749},
  number={1},
  pages={41},
  year={2012},
  publisher={IOP Publishing}
}

@article{richards2001photometric,
  title={Photometric redshifts of quasars},
  author={Richards, Gordon T and Weinstein, Michael A and Schneider, Donald P and Fan, Xiaohui and Strauss, Michael A and Vanden Berk, Daniel E and Annis, James and Burles, Scott and Laubacher, Emily M and York, Donald G and others},
  journal={The Astronomical Journal},
  volume={122},
  number={3},
  pages={1151},
  year={2001},
  publisher={IOP Publishing}
}

@article{fan1999high,
  title={High-Redshift Quasars Found in Sloan Digital SkySurvey CommissioningData},
  author={Fan, Xiaohui and Strauss, Michael A and Schneider, Donald P and Gunn, James E and Lupton, Robert H and Yanny, Brian and Anderson, Scott F and Anderson Jr, John E and Annis, James and Bahcall, Neta A and others},
  journal={The Astronomical Journal},
  volume={118},
  number={1},
  pages={1},
  year={1999},
  publisher={IOP Publishing}
}

@article{brescia2013photometric,
  title={Photometric redshifts for quasars in multi-band surveys},
  author={Brescia, Massimo and Cavuoti, STEFANO and D'Abrusco, R and Longo, G and Mercurio, AMATA},
  journal={The Astrophysical Journal},
  volume={772},
  number={2},
  pages={140},
  year={2013},
  publisher={IOP Publishing}
}

\appendix
\section{Comparison of regression methods for photo-z estimation}
\label{sec.comparison of photo_z methods}

\begin{figure*}
\centering
\begin{subfigure}{0.48\textwidth}
\includegraphics[width=\textwidth]{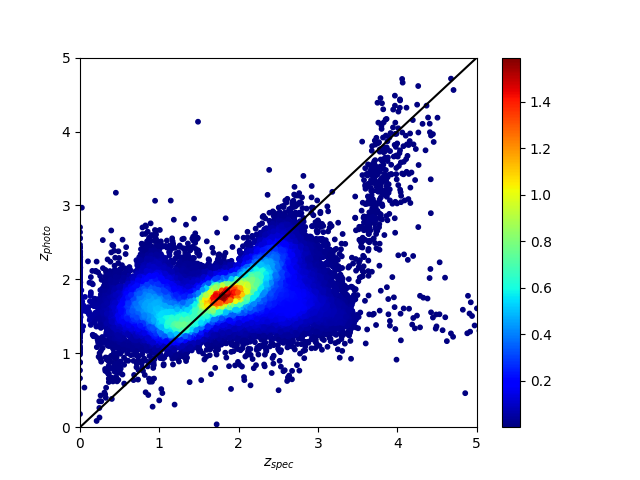}
\caption{Artificial Neural Network for magnitude-based predictors in {\tt ANNz}}
\label{fig.annz_ann_mags}
\end{subfigure}
\begin{subfigure}{0.48\textwidth}
\includegraphics[width=\textwidth]{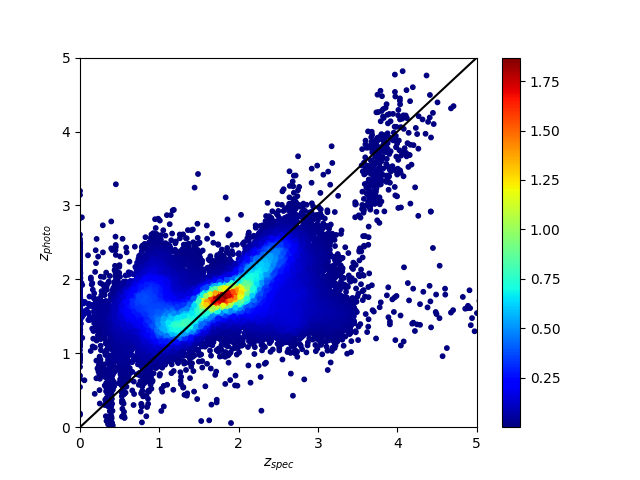}
\caption{Artificial Neural Network for color-based predictors in {\tt ANNz}}
\label{fig.annz_ann_colors}
\end{subfigure}
\vspace{1em}
\begin{subfigure}{0.48\textwidth}
\includegraphics[width=\textwidth]{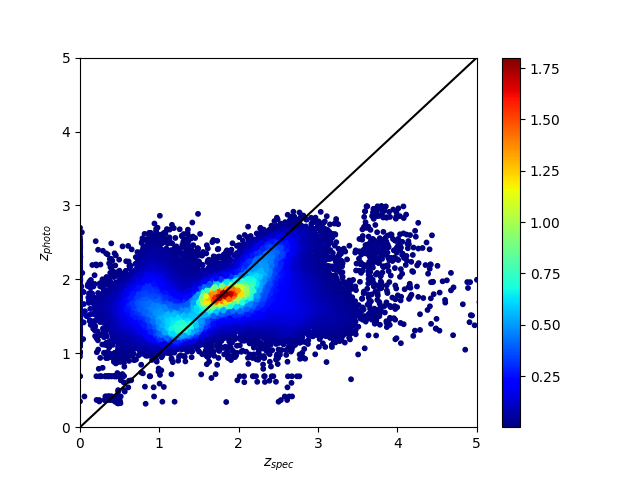}
\caption{Boosted Decision Tree for magnitude-based predictors in {\tt ANNz}}
\label{fig.annz_bdt_mags}
\end{subfigure}
\begin{subfigure}{0.48\textwidth}
\includegraphics[width=\textwidth]{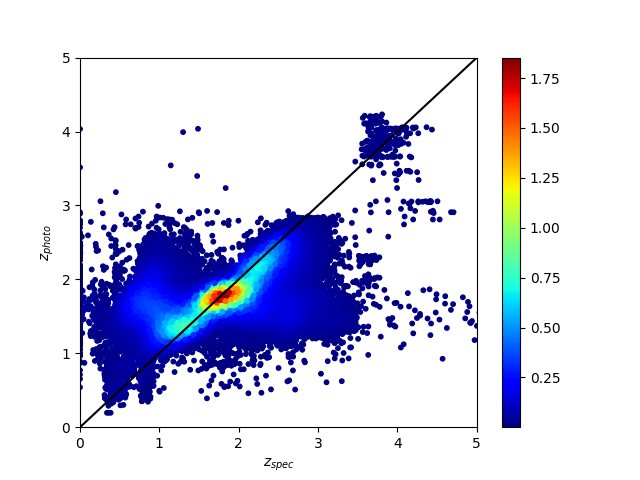}
\caption{Boosted Decision Tree for color-based predictors in {\tt ANNz}}
\label{fig.annz_bdt_colors}
\end{subfigure}
\caption{Photometric versus spectroscopic redshift for {\tt ANNz} implementations: (a) Artificial Neural Network implementation applied to magnitude parameters, (b) Artificial Neural Network implementation applied to color parameters, (c) Boosted Decision Tree implementation applied to magnitude parameters, (d) Boosted Decision Tree implementation applied to color parameters. The last one was chosen as better performance between the ANNz models.}
\label{fig.annz_photoz}
\end{figure*}

\begin{figure*}
\centering
\begin{subfigure}{0.48\textwidth}
\includegraphics[width=\textwidth]{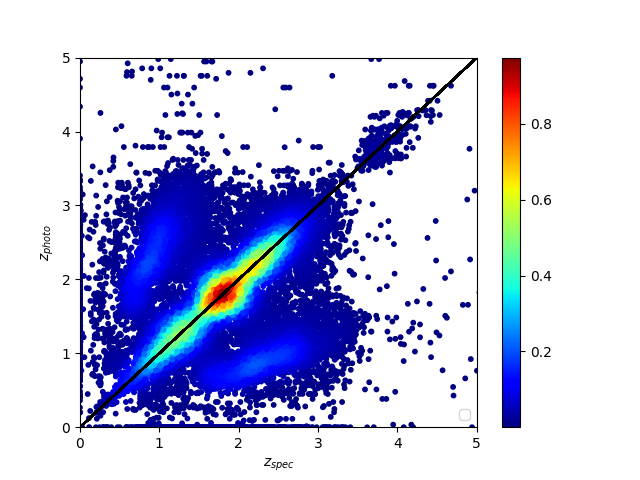}
\caption{Decision Tree Regressor.}
\label{fig.sk_dtr}
\end{subfigure}
\begin{subfigure}{0.48\textwidth}
\includegraphics[width=\textwidth]{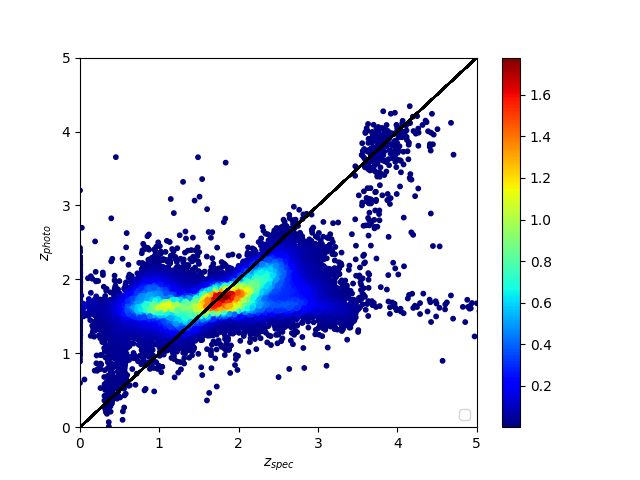}
\caption{Gradient Boosted Regressor.}
\label{fig.sk_gbr}
\end{subfigure}
\vspace{1em}
\begin{subfigure}{0.48\textwidth}
\includegraphics[width=\textwidth]{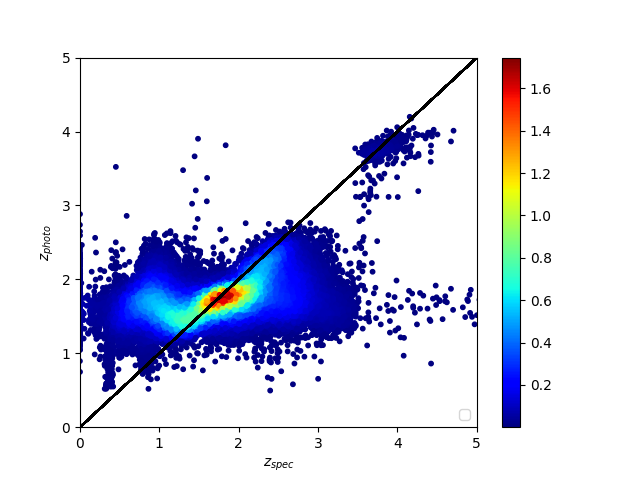}
\caption{K-Nearest Neighbors Regressor.}
\label{fig.sk_knr}
\end{subfigure}
\begin{subfigure}{0.48\textwidth}
\includegraphics[width=\textwidth]{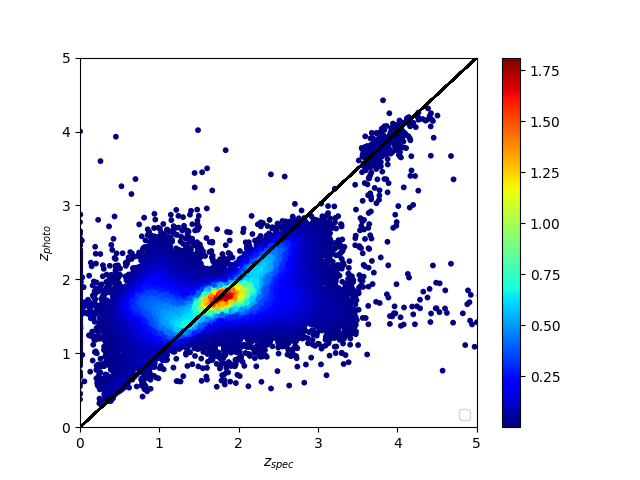}
\caption{Random Forest Regressor.}
\label{fig.sk_rfr}
\end{subfigure}
\caption{Photometric versus spectroscopic redshift for scikit-learn implementations. We evaluate the performance of different scikit-learn models: Decision Tree Regressor, Gradient Boosted Regressor, K-Nearest Neighbors Regressor and Random Forest Regressor respectively. The best performance was found with the Decision Tree Regressor, with large portion of the data aligned with the diagonal of the plot.}
\label{fig.sklearn_photoz}
\end{figure*}

To evaluate the impact of input features on model performance, we tested two distinct feature spaces in our photometric redshift estimation pipeline. The first, referred to as the magnitude space, consists of dereddened broad-band magnitudes from the DES photometric system: 
\lstinline{MAG_AUTO_G_DERED}, \lstinline{MAG_AUTO_R_DERED}, \lstinline{MAG_AUTO_I_DERED}, \lstinline{MAG_AUTO_Z_DERED}, and \lstinline{MAG_AUTO_Y_DERED}. The second, called the color space, includes four color indices—\lstinline{MAG_AUTO_G-R_DERED}, \lstinline{MAG_AUTO_R-I_DERED}, \lstinline{MAG_AUTO_I-Z_DERED}, and \lstinline{MAG_AUTO_Z-Y_DERED}—along with the dereddened r-band magnitude \lstinline{MAG_AUTO_R_DERED} and a morphological parameter, \lstinline{KRON_RADIUS}.

We systematically compared the performance of various machine learning models across both feature spaces using implementations from the {\tt ANNz} and {\tt scikit-learn} frameworks. This dual approach allowed us to assess the sensitivity of each model to the choice of input representation and to determine which combination of features and algorithms yielded the most accurate photometric redshifts.

For the overall behavior of the regression methods, we observe a prominent main cluster of objects where the photometric redshift estimates show small errors and closely track the spectroscopic redshifts, aligning well with the diagonal line. Additionally, two distinct off-diagonal clusters are consistently present; these correspond to objects for which the photometric redshift is not accurately recovered and are thus classified as outliers. Notably, we also identify a cluster near $z \approx 4 $, where photometric redshifts are well recovered, signifying the model's ability to handle high-redshift objects.

Upon evaluating the {\tt ANNz} models (Figure \ref{fig.annz_photoz}), we found that for some of the configurations, the $z \approx 4 $ cluster is not well formed or is dispersed. The best performing {\tt ANNz} model was the color-based Boosted Decision Tree (BDT), as shown in Figure \ref{fig.annz_bdt_colors}. This model exhibited a main cluster spanning a large redshift range and closely following the diagonal, a well-formed cluster in the $z \approx 4 $ region, and two distinct outlier groups.

Turning to the {\tt scikit-learn} implementations (Figure \ref{fig.sklearn_photoz}), we observed that for some models, the main cluster was not as well aligned across the diagonal as desired. The Decision Tree Regressor (DT), depicted in Figure \ref{fig.sk_dtr}, presented particularly good results in terms of its main cluster, which was very well formed and aligned across the diagonal, performing even better in this regard than the best {\tt ANNz} BDT model. The high-redshift cluster near $z \approx 4 $ was also well-formed. However, a limitation of the DT model was that its two outlier clusters were positioned further away from the main diagonal compared to other models.

Given the strengths of both the {\tt ANNz} color-based BDT model and the {\tt scikit-learn} Decision Tree Regressor (main cluster alignment), our subsequent idea was to combine the two best models to leverage their respective advantages. This combination aimed to achieve a well-defined main cluster while ensuring the two outlier clusters were not too far from the diagonal. The results of this combination are presented in Figure 5a of the main paper.

\end{document}